\documentclass[prx,aps,showpacs,twocolumn,preprintnumbers,
amsmath,amssymb,superscriptaddress,longbibliography]{revtex4-1}
\usepackage[english]{babel}
\usepackage{amsmath}
\usepackage{amssymb}
\usepackage{amsfonts}
\usepackage{physics}
\usepackage{graphicx}
\usepackage[colorlinks=True,linkcolor=red,citecolor=Red,urlcolor=Red]{hyperref}
\usepackage{bookmark}
\usepackage[dvipsnames]{xcolor} 
\usepackage{braket}
\usepackage{nicefrac}
\usepackage{bm}
\usepackage{bbm}
\usepackage{braket}
\usepackage{slashed}
\usepackage{mathdots}
\usepackage[normalem]{ulem}
\usepackage{array}
\usepackage{makecell}
\usepackage{amsthm}

\usepackage{cmap}
\usepackage{setspace}
\usepackage{amstext}			
\usepackage{amsfonts}
\usepackage{bbold}
\usepackage{graphicx}
\usepackage{subfigure}
\usepackage{wrapfig}
\usepackage{esint}
\usepackage{hyperref}
\usepackage{multirow}
\usepackage{printlen}
\usepackage{placeins}
\usepackage{comment}
\usepackage{bm}
\usepackage{tikz-cd}
\usepackage{tkz-berge}
\usepackage[boxsize=1em]{ytableau}
\usepackage[english]{babel}
\usepackage{fancyhdr}
\usepackage[left=1in, right = 1in, top = 1in, bottom = 1in]{geometry}

  \DeclareUnicodeCharacter{2212}{-}

\DeclareMathOperator{\diag}{diag}
\DeclareMathOperator{\im}{Im}

\DeclareMathOperator{\sign}{sign}

\newcommand{\BH}{\mathcal{H}}

\newcommand{\BP}{\mathcal{P}}
\newcommand{\BT}{\mathcal{T}}

\newcommand{\BC}{\mathcal{C}}
\newcommand{\BM}{\mathcal{M}}

\newcommand{\BL}{\mathcal{L}}

\newcommand{\BF}{\mathcal{F}}

\newtheorem{theorem}{Theorem}

\allowdisplaybreaks

\begin{document}

 \title{
Converting non-Hermitian degeneracies of any order:\\ Hierarchies of exceptional points and degeneracy manifolds
 }
 \author{Grigory A. Starkov}
  \affiliation{Institute for Theoretical Physics and Astrophysics,
University of W\"urzburg, D-97074 W\"urzburg, Germany}
\affiliation{W\"urzburg-Dresden Cluster of Excellence ctd.qmat, Germany}
\email{grigorii.starkov@uni-wuerzburg.de}
 \author{Sharareh Sayyad}
 \affiliation{Department of Mathematics and Statistics, Washington State University, \\Pullman, Washington 99164-3113, USA}
 \affiliation{Institute for Numerical and Applied Mathematics, Georg-August University of G\"ottingen, Bunsenstra\ss e 3-5, G\"ottingen 37073, Germany}
 \email{Sharareh.sayyad@wsu.edu}

\begin{abstract}
The emergence of various types of degeneracies plays a crucial role in optimizing and engineering different physical phenomena in non-Hermitian physics.
    In our work, we focus on the derogatory Exceptional Points (EPs), which are characterized by multiple Jordan blocks corresponding to the same eigenvalue. We demonstrate that, under certain infinitesimal perturbations, a derogatory EP can be converted into an EP of different structure without varying the total order of degeneracy. In particular, such conversion can increase the size of the largest Jordan block and, hence, the sensitivity of the eigenspectrum to parameter variation, which is an important feature for practical applications. 
    Furthermore, by analyzing all possible conversions, we introduce hierarchies of degeneracies of the same order that appear when perturbing non-Hermitian systems.
    We systematically explore hierarchies in the absence of any symmetry and when pseudo-Hermitian symmetry is present. Our study facilitates engineering various degeneracies of non-Hermitian systems, paving the way to extending the implications of non-Hermitian physics.
\end{abstract}

 \maketitle

 \section{Introduction}

Non-Hermiticity naturally occurs in open systems that exchange particles or energy with their surroundings~\cite{Ashida2020non, Wang_2021}. It also arises when we theoretically treat an isolated system as being composed of smaller subsystems; this is the central idea behind the description of self-thermalization in closed/isolated Hermitian systems. These effective non-Hermitian systems may exhibit a unique type of spectral degeneracy, called an Exceptional Point (EP)~\cite{Miri2019exceptional}, which has no counterpart in Hermitian systems. EPs present exciting practical applications, such as adiabatic state switching~\cite{Arkhipov2024, Wu-25, Bai-25, Laha-25, Qu-26} and enhanced sensitivity in quantum sensors~\cite{Khajavikhan-17, lai2019observation, kononchuk2022exceptional, Wang-24, Parto2025enhanced, Behrouzi-25, Chowdhury-25, Zheng-25, Wiersig-26}. 

Depending on the nature of non-Hermiticity, such as open or closed systems, or subsystems of isolated systems, various approaches have been proposed to model these systems. Examples include vectorized Louvillian superoperators~\cite{Tonielli2020, Sayyad2021, Starchl2022, Sayyad2023, Starchl2024, Marche2024} and non-Hermitian effective Hamiltonians~\cite{Jian2019, Kawabata2021, Sayyad2022c, Lado2023, Yoshida2024, Sayyad2024d, Legal2024} designed to capture the effects of low- or high-energy self-energies. In this work, we aim to explore any non-Hermitian models that can be represented in matrix form.

Every square matrix can be characterized by its Jordan normal form, a block-diagonal matrix representation composed of Jordan blocks, where a subset of these blocks represents any degenerate eigenvalue.
In Hermitian systems, all Jordan blocks are necessarily of size one. This merely allows for the emergence of non-defective degeneracies, also known as $m$-fold diabolical points, where $m$ denotes the multiplicity of the degenerate eigenvalue.
In contrast, non-Hermitian matrices may be non-diagonalizable, leading to Jordan blocks of size greater than one.
When a single Jordan block, with a size larger than one, represents an eigenvalue, the resulting degeneracy is referred to as a non-derogatory EP~\footnote{We note that in this case the geometric multiplicity of the eigenvalue is one.}.
This type of EP constitutes the primary focus of most studies on degeneracies in non-Hermitian systems.
In addition to non-derogatory EPs, non-Hermitian matrices may also host multiblock (derogatory) EPs, where a degenerate eigenvalue is associated with multiple Jordan blocks in the Jordan normal form~\cite{bid_PhysRevResearch, bid_2025, shiralieva2025}.
These relatively unexplored derogatory EPs naturally arise i) in the spectrum of the vectorized Liouvillian superoperator when the associated non-Hermitian Hamiltonian is tuned to an EP, whether non-derogatory or derogatory~\cite{shiralieva2025} and ii) in the composite systems~\cite{Wiersig2025}.

In our recent work~\cite{Sayyad2026}, we provided a systematic characterization of multiblock EPs under generic and selected non-generic perturbations. In the present work, we focus on the non-generic case and demonstrate that under certain perturbations, derogatory EPs can change their internal structure while maintaining the total order of the degeneracy. Such EP conversion can occur under arbitrarily small perturbations. Keeping track of all possible conversions starting from a particular EP reveals hierarchies of degeneracies.

Geometrically speaking, the idea of hierarchies is tied to the concept of the manifold of a family of matrices with the same Jordan normal form. For a (derogatory) EP with $r$ Jordan blocks of sizes $m_1\geqslant m_2\geqslant\dotsc \geqslant m_r$, the manifold $\BM_{m_1,m_2,\dotsc,m_r}$, that hosts a family of all $n\times n$ matrices, is obtained by similarity transformations with arbitrary invertible matrices $S$,
\begin{equation}
    \BM_{m_1,m_2,\dotsc,m_r} = \{S*J_{m_1,m_2,\dotsc,m_r}*S^{-1}|S\in GL_n(\mathbb C)\}.
\end{equation}
Here, $GL_n(\mathbb{C})$ is the general linear group of degree $n$ over $\mathbb{C}$, $m_i\geqslant 1$ sets the size of the $i$-th Jordan block, and $J_{m_1,m_2,\dotsc,m_r}$ denotes the corresponding Jordan normal form. Without loss of generality, we focus on a single degenerate eigenvalue and therefore assume $\sum_{i=1}^r m_i = n$. Considering two different EPs, say type-A and type-B, belonging to the manifolds ${\cal M}_{A}$ and ${\cal M}_{B}$ respectively, we explain in this work that the conversion of type-$A$ EP into a type-$B$ EP under an infinitesimal perturbation is possible if and only if the manifold $\BM_A$ lies in the boundary of the manifold $\BM_B$. As a result, determining the possible EP-conversions is equivalent to analyzing the hierarchy of degeneracy manifolds under closure.

The structure of the hierarchy depends on the symmetries imposed on the non-Hermitian system. In this work, we analyze both the generic case, where no symmetries are present, and the exemplary case of pseudo-Hermitian symmetry, demonstrating how symmetry constraints can prohibit certain types of EPs. The choice of pseudo-Hermitian symmetry is dictated by its close relation to (generalized) $\BP\BT$-symmetry, satisfied by the vectorized Liouvillian superoperator, as well as to the pseudo-chiral and charge-conjugate ($\BC\BP$) symmetries.

Some of the practical applications of EPs rely on their characteristic response to perturbations: for an $n$-th order non-derogatory EP, the eigenvalues typically split as the $n$ branches of the complex root $\Delta^{1/n}$, where $\Delta$ parametrizes the strength of the perturbation. Increasing the order of EPs makes them more singular in their vicinity, which is, for example, advantageous for quantum sensing. In the case of derogatory EPs, Jordan block of size $m_i$ splits as $\Delta^{1/m_i}$ under generic perturbations (see our recent paper~\cite{Sayyad2026} and Refs.~\cite{Lidskii-1966, Moro-2003}), i.e., it mimics a non-derogatory EP of order $m_i$.
Through EP conversion, the size of a Jordan block can increase, effectively raising the order of the EP. As a result, the EP-conversion phenomenon discussed in this work can be viewed as a novel mechanism for engineering higher-order EPs.

 The paper is structured as follows.
 To demonstrate the phenomenon of EP conversion and the notion of EP manifolds, we present several intuitive examples in Sec.~\ref{motivation}.
 In Sec.~\ref{GeometryN}, we analyze the hierarchy of EP manifolds under closure in the case of no symmetries. 
 We extend the hierarchy analysis to the case of the pseudo-Hermitian symmetry in Sec.~\ref{GeometryPSH}.
 Section~\ref{sec:applications} provides an analysis of the EP conversions in the eigenspectrum of the Liouvillian superoperator of a pseudo-Hermitian model.
Finally, we conclude this paper and discuss further possibilities for extending our findings in Sec.~\ref{sec:conclusion}.

\section{Motivational Examples}
To further motivate the idea of hierarchy of EP manifolds and demonstrate engineering of degeneracies using these hierarchies, we present illustrative examples in this section. 
\label{motivation}
\subsection{$2\times2$ Matrices}
\begin{figure}[t]
\includegraphics[width=0.98\columnwidth]{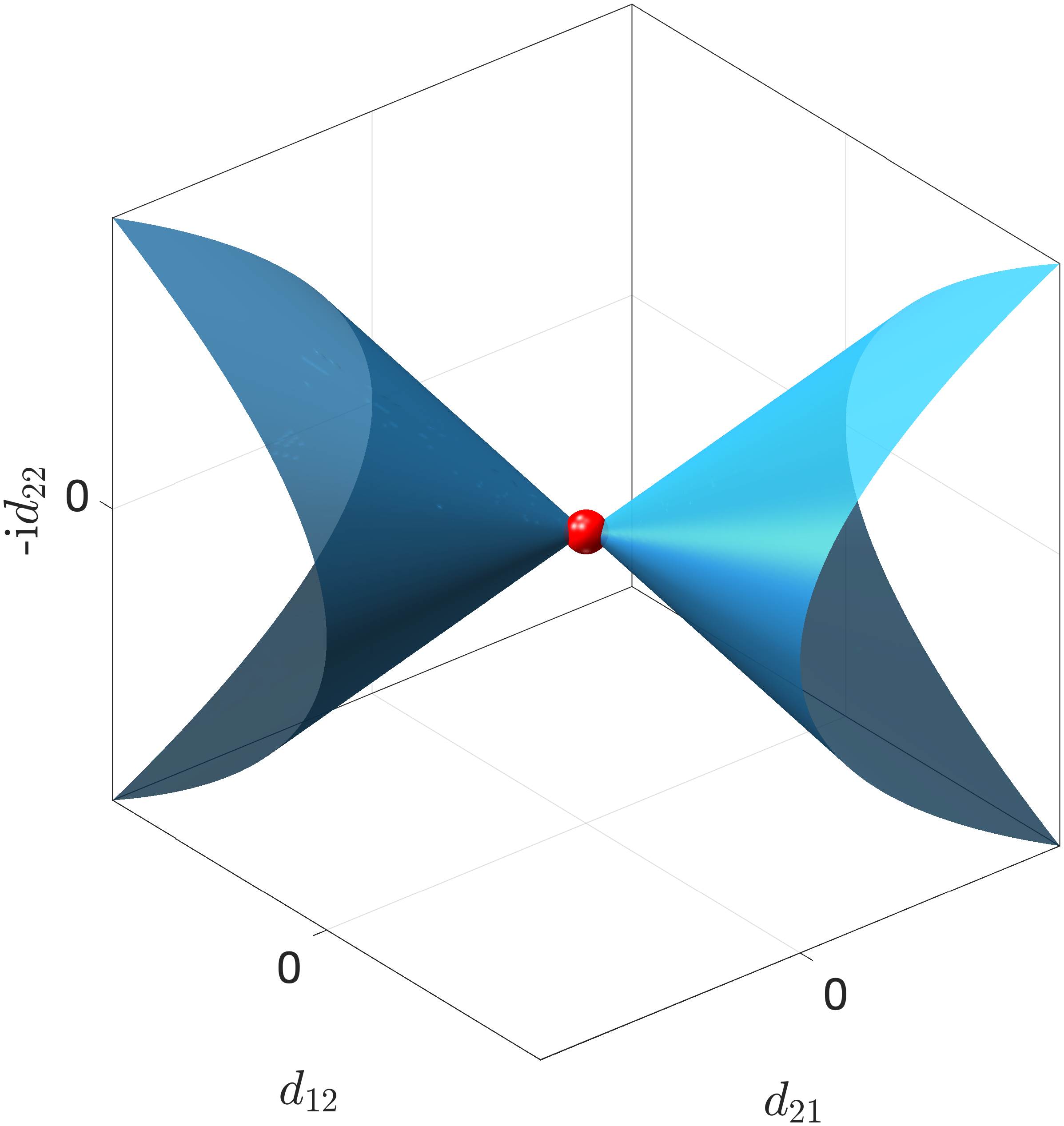}
\caption{Projection of the degeneracy surface $d_{22}^2+d_{12}d_{21}=0$ on the space $-id_{22},d_{12},d_{21}\in\mathbb R$. The red point marks the diabolical point.}
\label{fig:cone2d}
\end{figure}

Let us consider a generic $2\times2$ matrix $H$.
Without loss of generality, we assume that the trace of $H$ is set to zero~\footnote{We note that varying the trace merely adds a constant shift to the eigenvalues and does not affect the structure of degeneracies.}. The traceless matrix $H$ then reads
\begin{equation}
    H = \begin{pmatrix}
        -d_{22} & d_{12}\\
        d_{21} & d_{22}
    \end{pmatrix},\label{general22}
\end{equation}
Since we have three complex parameters~$\{d_{22}, d_{12}, d_{21}\}$, for fixed values of these matrix entries, we can view the matrix $H$ as a point in the $\mathbb C^3$ space.
The characteristic polynomial is
\begin{equation}
    \BF_\lambda = \det\left[\lambda \mathbb{1}_2 - H\right] = \lambda^2 - (d_{22}^2+d_{12}d_{21}),\label{ch-pol-general22}
\end{equation}
where $\mathbb 1_n$ is an $n\times n$ identity matrix.
The degeneracies occur when the condition $d_{22}^2+d_{12}d_{21} = 0$ is satisfied.
The polynomial in the left-hand side is homogeneous second order, therefore this condition defines a conical surface in $\mathbb C^3$: indeed, if $(d_{12}, d_{21},d_{22})$ is a solution, then $(td_{12}, td_{21},td_{22})$ also solves the degeneracy condition.
If we restrict the parameters $d_{12}$, $d_{21}$ to be purely real and $d_{22}$ to be purely imaginary, we can visualize the degeneracy condition, shown in Fig.~\ref{fig:cone2d}.

In the case of $2\times2$ matrices, we encounter two types of degeneracies: a diabolical point or a non-derogatory second-order EP. In the case of a diabolical point, a diagonalizable matrix with a degenerate eigenvalue is similar to
\begin{equation}
    \begin{pmatrix}
        \lambda & 0\\
        0 & \lambda
    \end{pmatrix}
\end{equation}
Requiring the matrix to be traceless results in setting $\lambda = 0$. The apex of the cone,  at $d_{12}=d_{21}=d_{22}=0$, corresponds to a diabolical point, while the remaining part of the cone represents second-order EPs. 

To demonstrate this explicitly, we note that all the points of the cone without the apex must correspond to matrices that are similar to the Jordan block of size two
\begin{equation}
    J_2 = \begin{pmatrix}
        0 & 1\\
        0 & 0
    \end{pmatrix}.
\end{equation}
Performing similarity transformations with arbitrary invertible matrices $S$,
\begin{equation}
    S = \begin{pmatrix}
        a & b\\
        c & d
    \end{pmatrix}, \qquad \text{with } ad-bc\neq 0,
\end{equation}
results in
\begin{equation}
    SJ_2S^{-1} = \frac{1}{ad-bc}\begin{pmatrix}
        -ac & a^2 \\
        -c^2 & ac
    \end{pmatrix}.
\end{equation}
This matrix automatically satisfies the degeneracy condition, and it is always non-zero: $a$ and $c$ can not be simultaneously zero, because otherwise $S$ would be non-invertible.

Perturbations along the cone leave the Jordan normal form and the eigenspectrum intact. So, to characterize the behavior of the eigenvalues, it is natural to focus on the perturbations that are orthogonal to the cone. This is the idea behind Arnold's canonical form for families of perturbed matrices~\cite{Arnold1971}. To put it in more precise terms, let us parametrize a perturbed second-order EP as
\begin{equation}
    H + \delta H = S\,(J_2 + \delta J)\,S^{-1},
\end{equation}
where $\delta J$ accounts for the perturbation of the Jordan normal form $J_2$.
The trivial perturbation along the cone can be obtained by varying $S$, denoted by $\delta S$,
\begin{align}
    \delta H_\mathrm{triv} &= \delta S  J_2 S^{-1} + S J_2 \delta S^{-1},\\
    &=S[S^{-1}\delta S, J_2]S^{-1},\\
    &= S \, \delta J_{\rm triv} \, S^{-1},
\end{align}
where we have used
\begin{equation}
    \delta(S\, S^{-1}) = \delta S \,S^{-1} +S\, \delta S^{-1}=0.
\end{equation}
Using this result, we can split the perturbation $\delta J$ into its trivial $\delta J_{\rm triv}= [\delta \tilde S, J_2]$ and non-trivial components:
\begin{equation}
    \delta J = \delta J_{\rm triv} + \delta J_\mathrm{n-triv},
\end{equation}
where $\delta \tilde S=S^{-1} \, \delta S$.
Note that this argument is also applicable to degeneracies in larger dimensions without any further modifications.

For an arbitrary $\delta \tilde S$ given by
\begin{equation}
    \delta \tilde S = \begin{pmatrix}
        \delta \tilde a& \delta \tilde b\\
        \delta \tilde c& \delta \tilde d
    \end{pmatrix},
\end{equation}
the commutator with Jordan block reads
\begin{equation}
    [\delta \tilde S, J_2] = \begin{pmatrix}
        \delta\tilde c & \delta\tilde a-\delta\tilde d\\
        0 & -\delta\tilde c
    \end{pmatrix}.
\end{equation}
As a result, for a perturbation
\begin{equation}
    \delta J= \begin{pmatrix}
        \delta J_{1,1} & \delta J_{1,2}\\
        \delta J_{2,1} & \delta J_{2,2}
    \end{pmatrix},
\end{equation}
all the matrix elements except $\delta J_{2,1}$ can be reabsorbed into the trivial perturbation. So, the general non-trivial perturbation looks like
\begin{equation}
    \delta J_\mathrm{n-triv} = \begin{pmatrix}
        0 & 0\\
        \delta J_{2,1} & 0
    \end{pmatrix},
\end{equation}
and the corresponding characteristic polynomial is
\begin{equation}
    \BF_\lambda = \det\left[\lambda \mathbb{1}_2 - (J_2+\delta J_\mathrm{n-triv})\right] = \lambda^2 - \delta J_{2,1}.
\end{equation}

We note that in the case of a second-order EP, any non-trivial perturbation lifts off degeneracy. This should be contrasted with the case of a diabolical point where any non-zero matrix~\eqref{general22} is a non-trivial perturbation in this case, and the characteristic polynomial of a perturbed diabolical point is given by Eq.~\eqref{ch-pol-general22}. However, in the case of perturbed diabolical points, even an infinitesimal perturbation satisfying the degeneracy condition $d_{22}^2+d_{12}d_{22} = 0$ keeps the characteristic polynomial unchanged, effectively transforming the diabolical point into a second-order EP.

\subsection{$3\times3$ Matrices}\label{sec:33}

A general traceless $3\times3$ matrix reads
\begin{equation}
    \begin{pmatrix}
 d_{11} & d_{12} & d_{13} \\
 d_{21} & -d_{11}-d_{33} & d_{23} \\
 d_{31} & d_{32} & d_{33}
\end{pmatrix},
\end{equation}
with the characteristic polynomial being~\cite{Sayyad2026}
\begin{equation}
{\cal F}_{\lambda} = \lambda^3-p\lambda-q.
\label{eq:charpolH111}
\end{equation}
Here, $p$ and $q$ are homogeneous polynomials of orders $2$ and $3$, respectively, given by
\begin{align}
   p =& d_{11}^2+d_{33} d_{11}+d_{33}^2 \nonumber
   \\ &+d_{12} d_{21}+d_{13} d_{31}+d_{23} d_{32}
    \label{ch-pol-p33}
    ,\\
    q =& d_{13} d_{31} d_{33}-d_{33}^2 d_{11}+d_{13} d_{31} d_{11} \nonumber \\
    &+d_{12} d_{23} d_{31}+d_{13} d_{21} d_{32}-d_{33} d_{11}^2\nonumber \\
   &-d_{12} d_{21} d_{33} -d_{23} d_{32} d_{11}.
      \label{ch-pol-q33}
\end{align}
Analogously to the case of $2\times2$ matrices, the condition of third-order degeneracy $p=q=0$ defines a conical surface in $\mathbb{C}^8$.

The zero matrix, which is the apex of this conical surface, corresponds to the tribolical point. Any perturbation of the tribolical point is non-trivial, and any infinitesimal perturbation satisfying $p=q=0$ leaves the characteristic polynomial invariant and transforms the tribolical point into an EP.

In the case of $3\times3$ matrices, in addition to the non-derogatory third-order EP, we also have type-$(2,1)$ EP, i.e., the EP corresponding to the Jordan normal form consisting of two blocks of sizes $2$ and $1$, respectively. The manifold of non-derogatory EPs is obtained by applying similarity transformations on the Jordan block of size $3$ as
\begin{align}
    M_3 &= \{S\, J_3\, S^{-1}|S\in GL_3 (\mathbb C)\},\\ J_3 &= \begin{pmatrix}
        0& 1 & 0\\
        0 & 0 & 1\\
        0 & 0 & 0\\
    \end{pmatrix}.
\end{align}
The manifold of type-$(2,1)$ EPs is obtained analogously from a direct sum of Jordan blocks of sizes $2$ and $1$
\begin{align}
    M_{2,1} &= \{S\, J_{2,1}\, S^{-1}|S\in GL_3(\mathbb C)\},\\ 
    J_{2,1} &= \begin{pmatrix}
        0 & 1 & 0\\
        0 & 0 & 0\\
        0 & 0 & 0
    \end{pmatrix}.
\end{align}
Both types of EPs correspond to a trivial characteristic polynomial of the form $\BF_\lambda = \lambda^3$. Consequently,  the union of two manifolds $M_3\cup M_{2,1}$ builds the conical surface $p=q=0$, excluding its apex. This stems from the fact that there are no other third-order degeneracy types, and the tribolical point corresponds to the apex. Hence, all the other points on the conical surface must be either non-derogatory or type-$(2,1)$ EPs.

Non-trivial perturbation of a non-derogatory third-order EP (orthogonal to the surface of $M_3$) is parametrized as~\cite{Sayyad2026}
\begin{align}
    H_3+\delta H_3^\mathrm{n-triv} &= S\, [J_3 + \delta J_3^\mathrm{n-triv}]\, S^{-1},\\ \delta J_3^\mathrm{n-triv} &= \begin{pmatrix}
        0 &  & 0\\
        0 & 0 & 0 \\
        \delta J_{3,1} & \delta J_{3,2} & 0
    \end{pmatrix},
\end{align}
and the resulting characteristic polynomial is
\begin{align}
    \BF_\lambda &= \det\left[\lambda \mathbb{1}_3 - (H_3+\delta H_3^\mathrm{n-triv})\right] ,\\
    &=\lambda^3 - \delta J_{3,2}\lambda - \delta J_{3,1}.
\end{align}
Parameters of the non-trivial perturbation directly correspond to the coefficients of the characteristic polynomial in this case, hence any non-trivial perturbation lifts degeneracy at least partially.

In comparison, non-trivially perturbed type-$(2,1)$ EPs, with perturbations orthogonal to the surface of $M_{2,1}$, are parametrized as
\begin{align}
    H_{2,1}+\delta H_{2,1}^\mathrm{n-triv} &= S\,[J_{2,1}+\delta J^{(2,1)}_\mathrm{n-triv}]\,S^{-1},\\
    \delta J^{(2,1)}_\mathrm{n-triv} &= \begin{pmatrix}
        0 & 0 & 0\\
        \delta J_{2,1} & -\delta J_{3,3} & \delta J_{2,3}\\
        \delta J_{3,1} & 0 & \delta J_{3,3}
    \end{pmatrix},
\end{align}
and the corresponding characteristic polynomial is
\begin{multline}
    \det\left[\lambda \mathbb{1}_3 - (H_{2,1}+\delta H_{2,1}^\mathrm{n-triv})\right]\\=
    \lambda^3 - (\delta J_{2,1}+\delta J_{3,3}^2)\lambda - (\delta J_{2,3}\delta J_{3,1}-\delta J_{2,1}\delta J_{3,3}).
\end{multline}
As we see, the codimension of $M_{2,1}$ (the number of parameters for non-trivial perturbations) is larger than the number of free parameters of the characteristic polynomial. As such, we can form an infinitesimally small non-trivial perturbation that does not lift the triple degeneracy by requiring
\begin{equation}
    \left\{ \begin{array}{l}
         \delta J_{2,1}+\delta J_{3,3}^2 =0,  \\
         \delta J_{2,3}\delta J_{3,1}-\delta J_{2,1}\delta J_{3,3}=0.
    \end{array}\right.\label{eq:sing21}
\end{equation}
A non-trivial perturbation must alter the type of degeneracy. And since an infinitesimal perturbation of a non-zero matrix can not result in a null matrix, the outcome must correspond to a non-derogatory third-order EP. Remarkably, we can transform a type-$(2,1)$ EP into a type-$3$ EP by a carefully chosen, arbitrarily small perturbation.

\begin{figure}
    \centering
    \includegraphics[width=0.98\columnwidth]{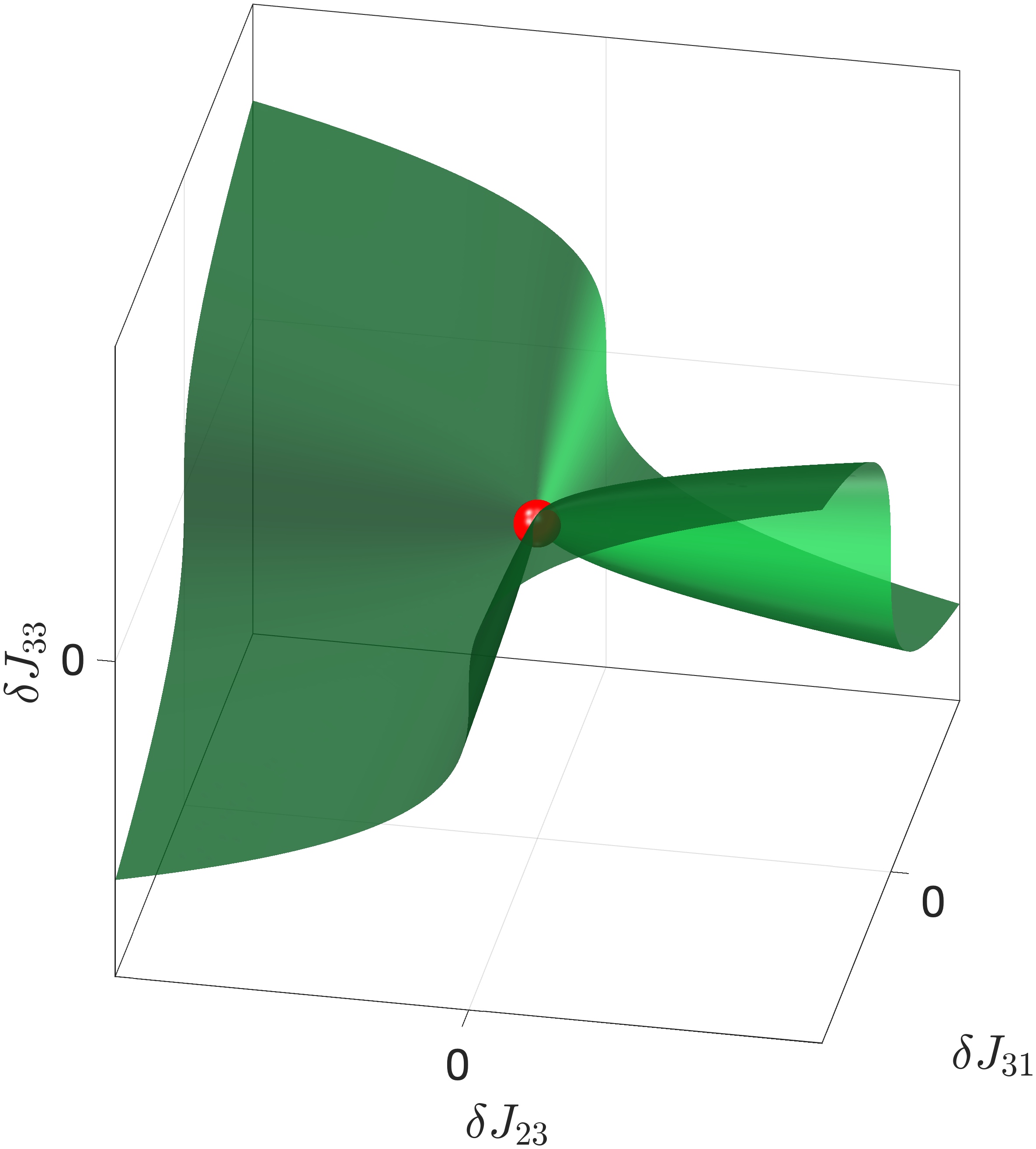}
    \caption{Manifold $M_3$ in the vicinity of a type-$(2,1)$ EP as described by Eq.~\eqref{eq:sing21}. All parameters have been restricted to be real, and $\delta J_{2,1}$ has been projected out as explained in the main text. The resulting equation of the surface is given in Eq.~\eqref{eq:sing21-projected}. The red point marks the type-$(2,1)$ EP.}
    \label{fig:sing21}
\end{figure}

To visualize how a type-$(2,1)$ EP attaches to the manifold $M_3$, we restrict all $\delta J_{i,j}$ to be real and project out $\delta J_{2,1}$. From the first equation in the system~\eqref{eq:sing21}, we get $\delta J_{21}=-\delta J_{33}^2$. Substituting this into the second equation gives the equation of the projected surface:
\begin{equation}
    \delta J_{2,3}\delta J_{3,1} + \delta J_{3,3}^3 = 0.\label{eq:sing21-projected}
\end{equation}
The resulting surface is shown in Fig.~\ref{fig:sing21}.

\subsection{$n\times n$ matrices}

In the case of an $n\times n$ matrix $H$, the corresponding characteristic polynomial is
\begin{equation}
    \BF_\lambda = \lambda^n - p_1\lambda^{n-1} - p_2\lambda^{n-2} -\dotsc-p_n,
\end{equation}
where the coefficients $p_k$ can be found using Faddeev-LeVierre method~\cite{Gantmakher1998, bid_PhysRevResearch}
\begin{equation}
    p_k = \frac{1}{k}\tr\left[A_k\right],\label{eq:p-eq}
\end{equation}
where
\begin{align}
A_1 &= H,\qquad p_1 = \tr\left[H\right],\label{eq:p-eq-init}
\\
    A_k &= A B_{k-1},\qquad B_k = A_k - p_k \mathbb{1}_{n\times n}.\label{eq:p-eq-recursion}
\end{align}
When we impose $H$ to be traceless, we get $p_1=0$.
Using Eqs.~\eqref{eq:p-eq}--\eqref{eq:p-eq-init}, it is straightforward to show that matrix elements of $A_k$ and $B_k$ as well as the coefficients $p_k$ are homogeneous polynomials of order $k$ in the matrix elements of $H$. As a result, the condition of the $n$-th order degeneracy
\begin{equation}
    p_k=0,\qquad k=2,3,\dotsc,n\label{eq:deg-condition},
\end{equation}
describes a conical surface in $\mathbb C^{n^2-1}$, where $(n^2-1)$ is the number of free parameters of a traceless $n\times n$ matrix.

For a non-derogatory $n$-th order EP, the non-trivially perturbed Jordan normal form is parametrized as~\cite{Arnold1971, Sayyad2022}
\begin{multline}
    J_n + \delta J_\mathrm{n-triv}^{(n)} = \\\begin{pmatrix}
        0 & 1 & 0 & \dotsc & 0 & 0\\
        0 & 0 & 1 & \dotsc & 0 & 0\\
        \vdots & \vdots & \vdots & \ddots & \vdots & \vdots\\
        0 & 0 & 0 & \dotsc & 1 & 0\\
        0 & 0 & 0 & \dotsc & 0 & 1\\
        \delta J_{n,1} & \delta J_{n,2} & \delta J_{n,3} & \dotsc &\delta J_{n, n-1} & 0
    \end{pmatrix}_{n\times n},
\end{multline}
with the corresponding characteristic polynomial
\begin{multline}
    \BF_\lambda = \lambda^n - \delta J_{n,n-1}\lambda^{n-2} - \delta J_{n,n-2}\lambda^{n-3} \dotsc \\\dotsc- \delta J_{n,2} \lambda - \delta J_{n,1}.
\end{multline}
Analogously to the non-derogatory EP2s and EP3s, the parameters of the non-trivial perturbation around a non-derogatory $n$-th order EP correspond to the coefficients of the characteristic polynomial. Therefore, a non-trivial perturbation generally changes the characteristic polynomial and lifts off the degeneracy.

For the derogatory EPs, the number of parameters for a non-trivial perturbation (which defines the codimension of the degeneracy manifold) is larger than for a non-derogatory EP~\cite{Arnold1971}, i.e., it is larger than the number of tunable coefficients of the characteristic polynomial. As a consequence, one can satisfy the conditions of degeneracy in Eq.~\eqref{eq:deg-condition} with non-zero non-trivial perturbations, which can be made arbitrarily small. Hence, {\it any} derogatory EP can be converted into a degeneracy of another type by a suitable infinitesimal perturbation.

In full analogy to the small order matrices, an $n$-bolical point corresponds to the zero $n\times n$ matrix. Any infinitesimal perturbation along the conical surface in Eq.~\eqref{eq:deg-condition} transforms it into an EP.

\subsection{Physical Example: non-Hermitian Lieb lattice}

Aside from the previous mathematically motivated examples, it is crucial to have an illustrative example with a more explicit physical interpretation. This is what we aim to present in the following.

In most experimentally-accessible systems, there is a limited number of control knobs that form the vector of parameters $\vec p$. The dependence of a Hamiltonian or a vectorized Liouvillian $H(\vec p)$ on $\vec p$ restricts the set of matrices realized by the system.
Geometrically speaking, the map $H(\vec p)$ defines a surface in the space of matrices. Intersections of this surface with the degeneracy manifolds define the physically observable degeneracy surfaces.

As a concrete example, let us consider the generalization of the lattice model used in Refs.~\cite{Bergholtz-21, bid_2025}
\begin{multline}
    H
    = \\\begin{pmatrix}
    0 & 1+e^{ik_y}+i\epsilon_1 & 0\\
    1+e^{-ik_y}+i\epsilon_2 & 0 & 1+e^{-ik_x} - i\epsilon_2\\
    0 & 1+e^{ik_x}-i\epsilon_1 & 0
    \end{pmatrix},\label{lieb-ham}
\end{multline}
where $\epsilon_1,\epsilon_2$ are non-Hermiticity parameters and $k_x$ and $k_y$ denote momentum along the $x$ and $y$ direction, respectively.
In the absence of non-Hermiticity, $\epsilon_1=\epsilon_2=0$, $H$ describes the Bloch Hamiltonian of the Lieb lattice with nearest-neighbor interactions, which hosts a tribolical point at $k_x=k_y=\pi$.

The eigenvalues of $H$ are $\varepsilon_{1,2,3} = 0,\pm\varepsilon_0$, where
\begin{multline}
    \varepsilon_0^2 = 2\left[2-\epsilon_1\epsilon_2 + (\cos{k_x}+\cos{k_y})+\vphantom{\frac{\epsilon_1-\epsilon_2}{2}}\right.\\+\frac{\epsilon_1-\epsilon_2}{2}(\sin{k_y}-\sin{k_x})+\\\left.\vphantom{\frac{\epsilon_1-\epsilon_2}{2}}+i(\epsilon_1+\epsilon_2)(\cos{k_y}-\cos{k_x})\right].
    \label{eq:eps0}
\end{multline}

\begin{figure*}
    \centering
    \includegraphics[width=\textwidth]{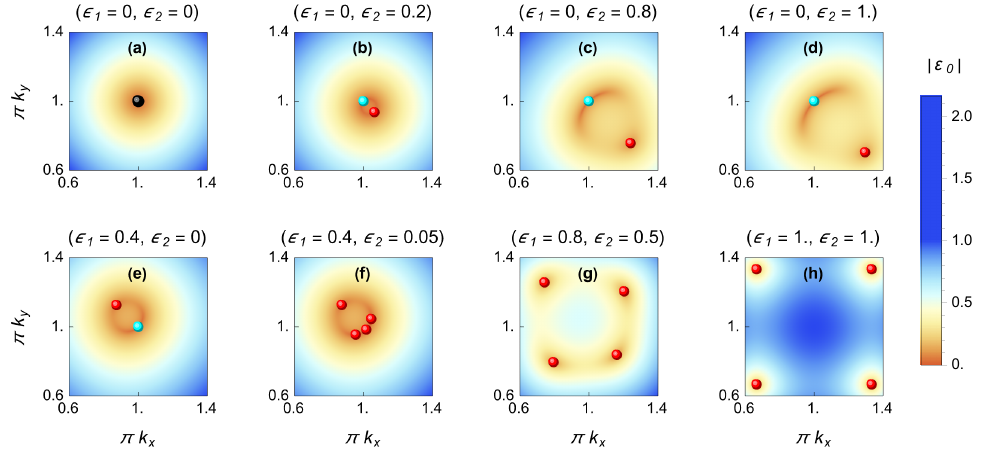}
    \caption{The absolute values of the spectrum of the model in Eq.~\eqref{lieb-ham}, given by Eq.~\eqref{eq:eps0}, at various values of $(\epsilon_1, \epsilon_2)$. Black points denote degeneracies at each set of parameters.
    The black point marks the tribolical point, cyan points show type-$(2,1)$ EPs, and red circles represent EP3s.
    }
    \label{fig:lieb_model}
\end{figure*}

Depending on the values of $\epsilon_1,\epsilon_2$, there can be up to four triple degeneracies in the $(k_x,k_y)$ plane as shown in Fig.~\ref{fig:lieb_model}. The first two points $(k_x,k_y)$ are present for
\begin{equation}
0\leqslant\epsilon_1\epsilon_2\leqslant4\label{region1},
\end{equation}
and have coordinates
\begin{equation}
(k_0,k_0), \text{ and } (-k_0,-k_0),
\end{equation}
where
\begin{equation}
k_0 = \arccos{\left(-1+\frac{\epsilon_1\epsilon_2}{2}\right)}.
\end{equation}
If the equality in Eq.~\eqref{region1} is satisfied, two points coalesce into one.

The other two triple degeneracies are present if
\begin{equation}
    [(\epsilon_1-1)(\epsilon_2-1)-1][(\epsilon_1+1)(\epsilon_2+1)-1]\leqslant0\label{region2},
\end{equation}
and have coordinates
\begin{align}
    k_0^\prime&=\arccos{\left(\frac{-1+\epsilon_1\epsilon_2/2}{\sqrt{1+(\epsilon_1-\epsilon_2)^2/4}}\right)},\\
    q &= \arctan{\left(\frac{\epsilon_1-\epsilon_2}{\sqrt{4+(\epsilon_1-\epsilon_2)^2}}\right)}.
\end{align}
Again, when the equality in Eq.~\eqref{region2} holds, the points merge into one.

Unless $\epsilon_1\epsilon_2=0$, all the triple degeneracies are non-derogatory EP3s, which can be determined from
\begin{equation}
    \rank{(\varepsilon\mathbb 1_3 - H)} = \rank{(H)} = 2.
\end{equation}
In other words, the system merely hosts a single linearly independent eigenvector~\footnote{This is a direct consequence of the rank-nullity theorem.}.

If $\epsilon_1\rightarrow0$ ($\epsilon_2\rightarrow0$), three of the EP3s coalesce into one point $(\pi,\pi)$, which is a type-$(2,1)$ EP; see panels b) and e) in Fig.~\ref{fig:lieb_model}. This result follows from the fact that $\rank{(\varepsilon\mathbb 1_3 - H)}=1$, which implies the existence of two eigenvectors.

As we see, the manifold of EP3s is two-dimensional, parametrized by $(\epsilon_1,\epsilon_2)$, and consists of several sheets. The type-$(2,1)$ EPs form two exceptional lines $\epsilon_{1/2}=0$ with a punctured point $\epsilon_1=\epsilon_2=0$, which corresponds to the tribolical point (shown by black point in Fig.~\ref{fig:lieb_model} (a)).
These two lines lie at the intersection of the sheets of the EP3 manifold\footnote{One should keep in mind that the lines themselves do not belong to the EP3 manifold. To get the intersection, we need to extend the sheets by their closure.}.

Overall, despite the form of the matrix is constrained by Eq.~\eqref{lieb-ham}, the picture we observe resembles the case of general $3\times3$ matrices we have discussed in Section~\ref{sec:33}: under infinitesimal perturbation, a tribolical point can be converted into EPs of any type, while a type-$(2,1)$ EP can be converted into EP3. The latter is typically destroyed by the perturbation.

\subsection{Formulation of the problem.}

All the examples we considered have one thing in common.
If we look at the Fig.~\ref{fig:cone2d}, we notice an important feature: the diabolical point (the apex of the cone) is the {\it boundary} of the manifold of second-order EPs (cone without its apex).
Analogously, analysis of Fig.~\ref{fig:sing21} leads us to the conclusion that a type-$(2,1)$ EP belongs to the boundary of the manifold $M_3$. This conclusion is equally true for the example of the non-Hermitian Lieb lattice, although the parameters of the Hamiltonian restricted the degeneracy manifolds.

Indeed, if a type-$B$ degeneracy can be obtained from a type-$A$ degeneracy by an arbitrarily small perturbation, then it means that any vicinity of a type-$A$ degeneracy, no matter how small, contains points of the $M_B$ degeneracy manifold, which is the mathematical definition of a limiting point. As such, the manifold $M_A$ of type-$A$ degeneracies must belong to the closure of the manifold of type-$B$ degeneracies. And the boundary is just the part of the closure different from the manifold itself.

This way, the physical problem ``Can we convert type-$A$ EP into a type-$B$ EP by an infinitesimal perturbation?'' is translated into the mathematical question ``Does the manifold $M_A$ belong to the closure of the manifold $M_B$?''

The latter question can be addressed by leveraging existing mathematical literature which we will elaborate on in the following sections.

 \section{Hierarchy of degeneracy manifolds in the case of no symmetries. \label{GeometryN}}

To study the relationship of the degeneracy manifolds under closure, let us first formalize the definition of these manifolds in the general case.

The type of degeneracy is uniquely determined by the corresponding Jordan normal form. So, for type-$(m_1,m_2,\dotsc,m_q)$ EP, the underlying matrix has the Jordan normal form
\begin{align}
    J_{m_1,m_2,\dotsc,m_q} = &J_{m_1}\oplus J_{m_2}\oplus\dotsc\oplus J_{m_q},\\
    J_m = &
    \begin{pmatrix}
        0 & 1 & 0 & \dotsc & 0 & 0\\
        0 & 0 & 1 & \dotsc & 0 & 0\\
        \vdots & \vdots & \vdots & \ddots & \vdots & \vdots\\
        0 & 0 & 0 & \dotsc & 1 & 0\\
        0 & 0 & 0 & \dotsc & 0 & 1\\
        0 & 0 & 0 & \dotsc &0 & 0
    \end{pmatrix}_{m\times m},
\end{align}
which is the direct sum of the Jordan blocks of sizes $m_i$.
To fix the ambiguity in the order of the blocks, the sizes are sorted in descending order $m_1\geq m_2\geq\dotsc \geq m_q$. We want to focus on the degeneracies of the same algebraic multiplicity $n$, so without loss of generality, we assume $\sum_{i=1}^q m_i = n$. The non-derogatory $n$-th order EP is a type-$(n)$ degeneracy, while $n$-bolical point is type-$(1,1,\dotsc,1)$ degeneracy, where $1$s are repeated $n$ times.

The manifold of type-$(m_1, m_2, \dotsc, m_q)$ degeneracies is obtained by acting on the corresponding Jordan normal form with similarity transformations
\begin{equation}
    M_{m_1,m_2,\dotsc,m_q} = \{SJ_{m_1,m_2,\dotsc,m_q}S^{-1}| S\in GL_n(\mathbb{C})\}.\label{sim-transform}
\end{equation}

One can show that the algebraic multiplicity is preserved under closure. Indeed, any $n$-th order degeneracy is characterized by a trivial characteristic polynomial $\BF_\lambda = \lambda^n$ and hence satisfies the system of polynomial equations in Eq.~\eqref{eq:deg-condition}, which is preserved under closure.
In addition to that, the closure is also invariant with respect to the similarity transformations~\footnote{A similarity transformation $A\rightarrow SAS^{-1}$ with fixed $S$ is a bounded linear operator on the space of matrices, hence it commutes with taking the limit.}. As a result, the boundary of the manifold $\overline M_{m_1,m_2,\dotsc,m_q}/M_{m_1,m_2,\dotsc,m_q}$ is always a union of one or several degeneracy manifolds of the same algebraic multiplicity $n$~\footnote{Note that $\overline{A}$ stands for the closure of set $A$.}.

To characterize the behavior of different degeneracy manifolds under closure, we can introduce the partial ``dominance'' order on the set of degeneracies of a given algebraic multiplicity~\cite{gerstenhaber_1959}.
We say that type--$(m_1,m_2,\dotsc,m_q)$ EP dominates type--$(m^\prime_1,m^\prime_2,\dotsc,m^\prime_{q^\prime})$ EP or
\begin{equation}
(m_1,m_2, \dotsc,m_q)\succ (m^\prime_1,m^\prime_2,\dotsc,m^\prime_{q^\prime}),
\end{equation}
if
\begin{equation}
M_{m^\prime_1,m^\prime_2,\dotsc,m^\prime_{q^\prime}}\subset\overline M_{m_1,m_2,\dotsc,m_q}.
\end{equation}
Here, we assume that $\sum_i{m_i} = \sum_i{m^\prime_i}=n$.

Reference~\cite{gerstenhaber_1959} proves that the condition of dominance is equivalent to
\begin{equation}
\rank{J_{m_1,m_2,\dotsc,m_q}^i}\geqslant\rank{J^i_{m^\prime_1,m^\prime_2,\dotsc,m^\prime_{q^\prime}}},
\label{dom-cond}
\end{equation}
for $i=1,\dotsc,n$. To determine the ranks systematically, the following notion would prove useful.

\begin{figure}
\ytableausetup{boxsize=1.5em}
\begin{ytableau}
1&2&3&4&5\\
6&7&8&9\\
10&11&12&13\\
14&15\\
16
\end{ytableau}
\caption{Example of a Young-diagram corresponding to the type--$(5,4,4,2,1)$ EP.}
\label{ydiag}
\end{figure}

We can represent the Jordan block partition corresponding to a type--$(m_1,m2_,\dotsc,m_q)$ degeneracy pictorially as a Young diagram with rows consisting of $m_1$, $m_2$, $\dotsc$, $m_q$ boxes, respectively (see Fig.~\ref{ydiag}). Moreover, we can associate each box with a vector of the Jordan basis, in which a matrix takes its Jordan normal form. When we act with $J_{m_1,m_2,\dotsc,m_q}$ on the vectors of the basis, we send each vector into the vector corresponding to the box left of it. Thus, for the example in Fig.~\ref{ydiag}, we for instance have
\begin{equation*}
\begin{array}{cc}
J_{5,4,4,2,1}e_5 = e_4, & J_{5,4,4,2,1}e_4 = e_3, \\
J_{5,4,4,2,1}e_9 = e_8, & J_{5,4,4,2,1}e_{12} = e_{11}.
\end{array}
\end{equation*}
 Here, the labels of the vectors correspond to the numbers inside the boxes in Fig.~\ref{ydiag}. The vectors corresponding to the leftmost boxes are sent to zero. For our example, these are the vectors $e_1$, $e_6$, $e_{10}$, $e_{14}$, $e_{16}$.

As we see, the $i$-th power of $H_{m_1,m_2,\dotsc,m_q}$ sends the first $i$ columns of a corresponding Young-diagram to zero. If we denote by $\hat m_{i}$ the column lengths, we can write explicitly~\cite{kraft_1982}
\begin{equation}
    \rank{J^i_{m_1,m_2,\dotsc,m_q}} = n - \sum_{j=1}^i \hat m_j.
\end{equation}
We assume here that $\hat m_j=0$ for $j>\hat q$, where $\hat q$ is the number of columns.
This way, the condition of dominance in Eq.~\eqref{dom-cond} takes the form
\begin{multline}
(m_1,m2_,\dotsc,m_q)\succ (m^\prime_1,m^\prime_2,\dotsc,m^\prime_{q^\prime})\Leftrightarrow\\\Leftrightarrow\sum_{j=1}^i \hat m_j\leqslant \sum_{j=1}^i\hat m^\prime_j,\quad i=1,2,\dotsc,\max\{\hat q,\hat q^\prime\}.\label{dominance}
\end{multline}

\ytableausetup{boxsize=0.6em}
\begin{figure}
\begin{tikzcd}
\ydiagram{2}\arrow{d}{}\\
\ydiagram{1,1}
\end{tikzcd}
\caption{Hierarchy of EPs with algebraic multiplicity $n=2$ consisting of partitions $(2)$ and $(1,1)$ from top to bottom.}
\label{eph2}
\end{figure}

\begin{figure}
\begin{tikzcd}
\ydiagram{3}\arrow{d}{}\\
\ydiagram{2,1}\arrow{d}{}\\
\ydiagram{1,1,1}
\end{tikzcd}
\caption{Hierarchy of EPs with algebraic multiplicity $n=3$ consisting of $(3)$, $(2,1)$ and $(1,1,1)$ from top to bottom.}
\label{eph3}
\end{figure}

\begin{figure}
\begin{tikzcd}
\ydiagram{4}\arrow{d}{}\\
\ydiagram{3,1}\arrow{d}{}\\
\ydiagram{2,2}\arrow{d}{}\\
\ydiagram{2,1,1}\arrow{d}{}\\
\ydiagram{1,1,1,1}
\end{tikzcd}
\caption{Hierarchy of EPs with algebraic multiplicity $n=4$ consisting of partitions $(4)$, $(3,1)$, $(2,2)$, $(2,1,1)$ and $(1,1,1,1)$ from top to bottom.}
\label{eph4}
\end{figure}

\begin{figure}
\begin{tikzcd}
\ydiagram{5}\arrow{d}{}\\
\ydiagram{4,1}\arrow{d}{}\\
\ydiagram{3,2}\arrow{d}{}\\
\ydiagram{3,1,1}\arrow{d}{}\\
\ydiagram{2,2,1}\arrow{d}{}\\
\ydiagram{2,1,1,1}\arrow{d}{}\\
\ydiagram{1,1,1,1,1}
\end{tikzcd}
\caption{Hierarchy of EPs with algebraic multiplicity $n=5$ consisting of partitions $(5)$, $(4,1)$, $(3,2)$, $(3,1,1)$, $(2,2,1)$, $(2,1,1,1)$ and $(1,1,1,1,1)$ from top to bottom.}
\label{eph5}
\end{figure}

\begin{figure}
\begin{tikzcd}
&\ydiagram{6}\arrow[Red]{d}{}&\\
&\ydiagram{5,1}\arrow[Red]{d}{}&\\
&\ydiagram{4,2}\arrow[Red]{dl}{}\arrow{dr}{}&\\
\ydiagram{4,1,1}\arrow[Red]{dr}{}&&\ydiagram{3,3}\arrow{dl}{}\\
&\ydiagram{3,2,1}\arrow[Red]{dl}{}\arrow{dr}{}&\\
\ydiagram{3,1,1,1}\arrow[Red]{dr}{}&&\ydiagram{2,2,2}\arrow{dl}{}\\
&\ydiagram{2,2,1,1}\arrow[Red]{d}{}&\\
&\ydiagram{2,1,1,1,1}\arrow{d}{}&\\
&\ydiagram{1,1,1,1,1,1}&
\end{tikzcd}
\caption{Hierarchy of EPs with algebraic multiplicity $n=6$ consisting of partitions $(6)$, $(5,1)$, $(4,2)$, ( $(4,1,1)$ and $(3,3)$), $(3,2,1)$, ($(3,1,1,1)$ and $(2,2,2)$), $(2,2,1,1)$, $(2,1,1,1,1)$ and $(1,1,1,1,1,1)$ from top to bottom. Red arrows mark the algebraic self-similar hierarchy of EPs constructed from Fig.~\ref{eph5}. }
\label{eph6}
\end{figure}

Using Eq.~\eqref{dominance}, we can arrange all the Young-diagrams, e.g., EPs, with a fixed algebraic multiplicity in a hierarchy according to their ``dominance''~\footnote{These structures are also known as the Young diagram lattices on the partition of an integer~\cite{Latapy2009}.}.
We have developed a \href{https://github.com/say-yas/Hierarchy_NH_degeneracies}{Mathematica notebook} and a \href{https://github.com/Gregstrq/DominanceHierarchy.jl}{Julia package} to facilitate finding these hierarchies.
The corresponding hierarchies for algebraic multiplicities $n=2,3,4,5,6,7$ are displayed in Figs.~\ref{eph2}, \ref{eph3}, \ref{eph4}, \ref{eph5}, and \ref{eph6}, respectively (see also Ref.~\cite{kraft_1982}). The degeneracies are labeled by their respective Young diagrams. The arrows represent the dominance relation: if we can get from a type--$A$ degeneracy to a type--$B$ degeneracy by moving down along the arrows, then $A\succ B$. And vice versa, if we can get from type--$B$ degeneracy to the type--$A$ degeneracy by moving up against the arrows, then we can get type--$A$ degeneracy by an infinitesimal perturbation of type--$B$ degeneracy.

For example, in the case of $n=2$, there are only two possibilities: type--$(2)$ EPs (the usual second-order EPs) or type--$(1,1)$ non-defective degeneracies (diabolical point). In agreement with what we have already mentioned, $(2)\succ(1,1)$.
In the case of $n=3$, there are three possibilities: type--$(3)$ EPs (non-derogatory third-order EPs), type--$(2,1)$ EP and type--$(1,1,1)$ degeneracy (tribolical point). Their ordering is $(3)\succ(2,1)\succ(1,1,1)$.
This agrees with our conclusion of the previous section that one can get a type--$(3)$ EP by an infinitesimal perturbation of type--$(2,1)$ EP.

Starting with $n=6$, we may finally notice that the ``dominance'' is only a partial order on the set of degeneracies. For example, neither $(4,1,1)\nsucc(3,3)$ nor $(3,3)\nsucc(4,1,1)$; See the 4th row from the top in Fig.~\ref{eph6}. As such, we can get neither type--$(3,3)$ by an infinitesimal perturbation of type--$(4,1,1)$ EP nor vice versa. On the contrary, $(4,1)\succ(4,1,1),(3,3)$, so we can obtain a type--$(4,1)$ EP by an infinitesimal perturbation of either type--$(4,1,1)$ or type--$(3,3)$ EP.

Irrespective of $n$, the hierarchies share some common features~\cite{kraft_1982}: At the top of a hierarchy always sits a type--$(n)$ EP. The closure of the corresponding manifold $M_n$ contains all the other degeneracy manifolds for the same algebraic multiplicity. At the bottom of the hierarchy always sits an $n$-bolical point (type--$(1^n)$), which corresponds to the null matrix of size $n\times n$. We can demonstrate these points explicitly if we consider the action of similarity transformations with diagonal $S = \diag{(s_1, s_2,\dotsc, s_n)}$ on Jordan block of size $n$:
\begin{multline}
    S\, J_n \,S^{-1} = \\\begin{pmatrix}
        0 & s_1/s_2 & 0 & \dotsb & 0 & 0\\
        0 & 0 & s_2/s_3 & \dotsb & 0  & 0 \\
        \vdots & \vdots & \vdots & \ddots & \vdots &\vdots\\
        0 & 0 & 0 & \dotsb & s_{n-2}/s_{n-1} &0 \\
        0 & 0 & 0 & \dotsb & 0 & s_{n-1}/s_{n}\\
        0 & 0& 0 &\dotsc & 0 & 0
    \end{pmatrix}_{n\times n}.
\end{multline}
If we take the limit $s_i/s_{i+1}\rightarrow 0$ while keeping all the other ratios fixed, we can nullify the $i$-th elements of the first upper diagonal. Analogously, we can actually nullify any subset of the elements of the first upper diagonal, obtaining Jordan normal forms for all degeneracies of algebraic multiplicity $n$.

There are also algebraic self-similar structures between the hierarchies of EPs with algebraic multiplicity $n$ and their counterparts with algebraic multiplicity $n+1$~\cite{Latapy2009}. To find the self-similar sublattices in hierarchies with $n+1$, one attaches one box to the first column of all Young diagrams in hierarchies with algebraic multiplicity $n$. As a result, a new hierarchy consisting of partitions from $n+1$ to $2, 1, \ldots,1$, with the count of ones being $n-1$; see Young diagrams connected by red arrows in Fig.~\ref{eph6}. For algebraic multiplicity $ n\leq 5$, the self-similar hierarchies comprise all partitions of $n$ except the last one associated with type--$(1^n)$ degeneracies. For $n >5$, the subgraph for each hierarchy consists of branches that are not present in the self-similar hierarchies.

\section{Hierarchy of degeneracy manifolds in the case of pseudo-Hermitian symmetry}
\label{GeometryPSH}

\subsection{Classification of degeneracies in pseudo-Hermitian systems}\label{sec:psh_1}

In the presence of symmetries, both allowed matrices and their perturbations are restricted, which complicates the picture. For example, let us say we want to describe a manifold corresponding to a specific degeneracy type. While the different points on this manifold are still related by the similarity transformation~\eqref{sim-transform}, $S$ can not be an arbitrary invertible operator since it has to preserve the symmetry. Moreover, we might need to rethink the classification of possible degeneracies itself.

If we pose the same question ``which other types of degeneracy can be accessed by an infinitesimal perturbation of a given degeneracy type?" the answer would depend on the symmetry under consideration and thus needs to be discussed on a case-by-case basis.
We will focus on the case of pseudo-Hermitian symmetry, which is closely related to the case of the $\BP\BT$-symmetry~\cite{mostafazadeh2002pseudo, mostafazadeh2, mostafazadeh3, mostafazadeh2010, ZhangQinXiao-20, Ashida2020non}. The latter one is relevant for the analysis of the spectrum of the Liouvillian superoperator, which is known to satisfy a generalized $\BP\BT$-symmetry~\cite{Sa_2023}. In addition to that, the transformation $H\rightarrow iH$ maps the PT- and pseudo-Hermitian-symmetric systems onto the CP- and pseudo-chiral-symmetric ones, respectively.

An operator $H$ is said to be pseudo-Hermitian if
\begin{equation}
    H = \eta^{-1} H^\dagger \eta, \label{pH-def}
\end{equation}
where $\eta$ is an invertible Hermitian operator called pseudo-metric.
If $\eta$ does not depend on the parameters of $H$, then we have a pseudo-Hermitian symmetry.

The pseudo-metric operator defines a nondegenerate scalar product in terms of the standard scalar product of the Hilbert space
\begin{equation}
    (\psi,\chi) = \langle\psi|\eta|\chi\rangle.
\end{equation}
The right-hand-side of Eq.~\eqref{pH-def} defines the adjoint operator with respect to this scalar product
\begin{equation}
    (\psi,A\chi) = (A^{[\dagger]}\psi,\chi),\qquad \mathrm{with}\qquad A^{[\dagger]} = \eta^{-1} A^\dagger \eta.
\end{equation}
Correspondingly, the pseudo-Hermitian symmetry means that the operator is self-adjoint with respect to the scalar product defined by the pseudo-metric.

Applying a similarity transformation to Eq.~\eqref{pH-def} results in
\begin{multline}
  H'=  S H S^{-1} =\\ \left(\left(S^{-1}\right)^{\dagger}\eta S^{-1}\right)^{-1} \left(S H S^{-1}\right)^{\dagger} \left(\left(S^{-1}\right)^\dagger\eta S^{-1}\right).
\end{multline}
As we see, the transformed Hamiltonian $H^\prime$ still satisfies a pseudo-Hermitian symmetry, albeit with respect to a transformed pseudo-metric $\eta^\prime = \left(S^{-1}\right)^\dagger\eta S^{-1}$. If we now require that the similarity transformation preserves the pseudo-Hermitian symmetry, this results in the condition
\begin{equation}
    \eta^\prime=\eta \Leftrightarrow S^\dagger\eta S = \eta. \label{sym-preserve}
\end{equation}
It is equivalent to saying that $(S\psi,S\chi) = (\psi,\chi)$, {\it i.e.}\/, the transformation preserves the scalar product associated with pseudo-metric.

With a suitable basis change $\eta = S_0^\dagger \eta_{p,q} S_0$, invertible Hermitian $\eta$ can be transformed into the diagonal form~\footnote{A Hermitian operator $\eta$ can be diagonalized by a unitary transformation: $\eta = U^\dagger D U$, where $U$ is an invertible matrix and $D = \diag{(\varepsilon_1, \varepsilon_2,...)}$ with real eigenvalues $\varepsilon_i$. 
Without the loss of generality let's assume that the eigenvalues are ordered $\varepsilon_i\geqslant \varepsilon_i+1$, first $p$ eigenvalues are positive and the rest $q$ eigenvalues are negative. Then $\eta = S_0^\dagger \eta_{p,q}S_0$ with $S_0 = |D|^{1/2}U$, where $|D|^{1/2} = \diag{(\sqrt{|\varepsilon_1|}, \sqrt{|\varepsilon_2|},...)}$}
\begin{equation}
  \eta_{p,q} = \begin{pmatrix}
        \mathbb{1}_p & 0\\
        0 & -\mathbb{1}_q\\
    \end{pmatrix},\label{pmetric-sform}
\end{equation}
where $p$ and $q$ are total numbers of positive and negative eigenvalues of $\eta$, respectively.
Substituting transformed $\eta$ into Eq.~\eqref{sym-preserve} gives
\begin{equation}
     (S_0SS_0^{-1})^\dagger \eta_{p,q} \left(S_0 SS_0^{-1}\right) = \eta_{p,q},
\end{equation}
which means that $S_0 SS_0^{-1}$ belongs to the indefinite Unitary group $U(p,q)$ and
\begin{equation}
    S = S_0^{-1} W S_{0},\qquad W\in U(p,q).\label{sym-group}
\end{equation}
The similarity transformation~\eqref{sym-group} establishes isomorphism between the group preserving the pseudo-Hermitian symmetry and the indefinite Unitary group $U(p,q)$, where $p$ and $q$ are determined by the signature of the pseudo-metric operator~\cite{Mostafazadeh2004}.

We can still characterize different degeneracies of the operator $H$ using its Jordan normal form. However, the presence of pseudo-Hermitian symmetry actually restricts the possible Jordan form structures, and the classification of degeneracies needs to be refined. First of all, this type of symmetry makes the coefficients of the characteristic polynomial strictly real, which means that all the eigenvalues come in complex conjugate pairs. Moreover, the Jordan blocks themselves come in complex conjugate pairs~\cite{Gohberg2006}: for every Jordan block $J_k(\lambda)$ of size $k$ in the Jordan normal form, there is a block $J_k(\bar\lambda)$. Secondly, there is a certain freedom in the choice of the basis of a linear subspace corresponding to a particular Jordan block. This freedom can be used to bring the pseudo-metric operator to a particular simple form simultaneously with $H$. To do so, we employ the following theorem~\footnote{ See Theorem 5.1.1 in Ref.\cite{Gohberg2006}.}.
\begin{theorem} \label{theorem1}
Let $H\in C^{n\times n}$ be pseudo-Hermitian~(self-adjoint) operator with respect to the Hermitian invertible $\eta\in {\mathbb C}^{n\times n}$. Then, there is an invertible $T\in {\mathbb C}^{n\times n}$ such that
\begin{equation}
    H = T^{-1} J T,\qquad \eta = T^\dagger PT,\label{th-transform}
\end{equation}
where
\begin{align}
    J =& J_{m_1}(\lambda_1)\oplus \dotsc\oplus J_{m_\alpha}(\lambda_\alpha)
    \nonumber \\
    &
    \oplus \tilde J_{m_{\alpha+1}}(\lambda_{\alpha+1})\oplus\dotsc \oplus \tilde J_{m_\beta}(\lambda_{\beta}),
    \label{jordan-pH}
\end{align}
is the Jordan normal form for $H$ with $J_{n}(\lambda)=J_n+\lambda \mathbb{1}_n$ being the standard Jordan matrix of size $n$ corresponding to the eigenvalue $\lambda$ and $\tilde J_{n}(\lambda) = J_n(\lambda)\oplus J_n(\bar\lambda)$. Here, $\lambda_1,\dotsc,\lambda_\alpha$ are real eigenvalues of $H$, and $\lambda_{\alpha+1},\dotsc,\lambda_{\beta}$ are complex eigenvalues with positive imaginary parts, and $m_{i}$s set the size of Jordan block associated with eigenvalue $\lambda_{i}$. Eigenvalues can repeat themselves. Also,
\begin{equation}
    P = \varepsilon_1 P_{m_1}\oplus \dotsc \varepsilon_\alpha P_{m_\alpha}\oplus P_{2m_{\alpha+1}}\oplus\dotsc P_{2m_\beta},\label{pmetric-expand}
\end{equation}
where $\varepsilon_i=\pm1$ and $P_l$ is an antidiagonal matrix~\footnote{This matrix is also known as the standard involutory permutation~(sip) matrix, or the reversal matrix~\cite{horn2012matrix}.} of size $l\times l$ given by
\begin{equation}
    P_l = \begin{pmatrix}
        0 & 0 & \dotsb & 0 & 1\\
        0 & 0 & \vdots & 1 & 0\\
        \vdots & \vdots & \iddots & \vdots & \vdots\\
        0 & 1 & \dotsb & 0 & 0\\
        1 & 0 & \dotsb & 0 & 0
    \end{pmatrix}.
\end{equation}
The signs $\varepsilon_i$ are defined up to a permutation of the Jordan blocks of the same size corresponding to the same eigenvalue.
\label{theo}
\end{theorem}

Each term in Eq.~\eqref{pmetric-expand} describes the restriction of the pseudo-metric operator to the linear subspace corresponding to the respective term in Eq.~\eqref{jordan-pH}. The matrix $P_l$ is diagonalizable and has eigenvalues $\pm1$. If $l$ is even, there are equal numbers of eigenvectors corresponding to $+1$ and $-1$. If $l$ is odd, the number of eigenvectors corresponding to $+1$ is one larger. The transformation in Eq.~\eqref{th-transform} does not change the number of positive $p$ and negative $q$ eigenvalues of the pseudometric, therefore the signatures of $\eta$ and $P$ are identical
\begin{equation}
    p-q = \sum_{i=1}^{\alpha} \frac{1 - (-1)^{m_i}}{2}\varepsilon_i.\label{signature-cond}
\end{equation}
On the right-hand side, we summed up the signatures of different Jordan blocks. 
The even-sized Jordan blocks do not contribute to the signature, and thus, all the blocks corresponding to the complex conjugate pairs are excluded from the sum.
Equation~\eqref{signature-cond} limits the degrees and the numbers of Exceptional points available in the system. For example, if the pseudometric is positive (negative) definite, then the system cannot host any exceptional points; only non-defective degeneracies may find room to emerge. Moreover, a non-derogatory exceptional point of maximal degree $n=p+q$ can be possible only when $p-q = 0,\pm1$.

As the parameters of the Hamiltonian are varied, real eigenvalues can collide and become complex. Analysis of the form~\eqref{pmetric-expand} constrains for which eigenvalues it can happen. For simplicity, let us consider collision of several non-degenerate eigenvalues $\lambda_{i_1},\lambda_{i_2}, \dotsc,\lambda_{i_m}$, that form a non-degenerate EP as a result.
The eigensubspaces of the non-degenerate eigenvalues are one-dimensional, and the restriction of the pseudometric to such a subspace is simply a scalar:
\begin{equation}
    \varepsilon_{i_k}P_{i_k} = \varepsilon_{i_k} = \sign{(v_{i_k},v_{i_k})},
\end{equation}
where $v_{i_k}$ is the corresponding eigenvector.
For pseudo-Hermitian $H$, projection of the pseudometric on an eigensubspace or sum thereof has signature which is stable against small perturbations of $H$~\footnote{For example, it follows from Theorem 9.1.1 in Ref.~\cite{Gohberg2006}}. As a first consequence, we see that the signs $\varepsilon_{i_k}$ do not change unless the corresponding eigenvalue goes through an EP. Secondly, when the eigenvalues collide and form a non-degenerate EP, the restriction of the pseudo-metric to the corresponding linear subspace would have eigenvalues $(\varepsilon_{i_1},\varepsilon_{i_2},\dotsc,\varepsilon_{i_m})$.
If only ``positive" or only ``negative" eigenvalues collide, we would come in contradiction with Eq.~\eqref{pmetric-expand}. So, to form an exceptional point eigenvalues with different $\varepsilon_i$ have to collide. This phenomenon is known as Krein collision in the context of linear systems with periodic coefficients~\cite{Krein-1950, GelfandLidskii-1955} and has recently been discussed in the context of non-Hermitian physics~\cite{Melkani-23, StarkovFistulEremin-23, StarkovFistulEremin-23b, Starkov-24}.

\subsection{Hierarchy of degeneracies}

Let us look closely at the respective blocks $\tilde J_{m_{\alpha+i}}(\lambda_{\alpha+i})$ and $P_{2m_{\alpha+i}}$ in the Jordan normal form and transformed pseudo-metric corresponding to the same complex energy $\lambda_{\alpha+i}$. We can write them explicitly as
\begin{equation}
    \tilde J_{m_{\alpha+i}}(\lambda_{\alpha+i}) = \begin{pmatrix}
        J_{m_{\alpha+i}}(\lambda_{\alpha+i}) & 0_{m_{\alpha+i}\times m_{\alpha+i}}\\
        0_{m_{\alpha+i}\times m_{\alpha+i}} & J_{m_{\alpha+i}}(\bar\lambda_{\alpha+i})
    \end{pmatrix},
\end{equation}
\begin{equation}
    P_{2m_{\alpha+i}} = \begin{pmatrix}
        0_{m_{\alpha+i}\times m_{\alpha+i}} & P_{m_{\alpha+i}}\\
        P_{m_{\alpha+i}} & 0_{m_{\alpha+i}\times m_{\alpha+i}}
    \end{pmatrix}.\label{p-complex}
\end{equation}
When we restrict the pseudometric to the linear subspace corresponding to a single complex energy without its complex conjugate partner, the result is zero (diagonal blocks in Eq.~\eqref{p-complex}).
Consequently, the behavior of a degeneracy at a complex energy under perturbation is similar to the non-symmetric case of Sec.~\ref{GeometryN} as it corresponds to the trivial zero pseudo-metric~\footnote{There is no contradiction here. Pseudo-Hermitian symmetry is still satisfied in the sense that the corresponding degeneracy at the complex conjugate energy behaves symmetrically.}. Therefore, we will focus our attention on the degeneracies at real energies.

Without loss of generality, we assume that other energy levels have been projected out and focus on the single eigenvalue. We assume the pseudo-metric to be restricted to the corresponding subspace and choose the basis in which it takes the simple form~\eqref{pmetric-sform}. Since the actual value of the eigenvalue is not important, we set it to be zero and consider traceless pseudo-Hermitian matrices. For the above-mentioned choice of the basis, we can associate the space of all traceless pseudo-Hermitian matrices with the $su(p,q)$ Lie algebra:
\begin{equation}
    \eta_{p,q} H - H^\dagger \eta_{p,q}=0 \Rightarrow \eta_{p,q} (iH) + (iH)^\dagger \eta_{p,q}=0,
\end{equation}
with $(iH)\in su(p,q)$.

From Theorem~\ref{theorem1} it follows that every degeneracy is uniquely characterized by the Jordan decomposition $J_{m_1,m_2,\dotsc,m_r}$ and the set of signs $(\varepsilon_1,\varepsilon_2,\dotsc,\varepsilon_r)$, which together can be associated with a signed Young diagram~\cite{Djokovic2006, Djokovic1982, Collingwood2017}. In comparison to a simple Young diagram introduced in the previous section, each row here also carries alternating pluses and minuses. The right-most sign corresponds to the sign $\varepsilon_i$ in the decomposition of the pseudo-metric~\eqref{pmetric-expand}, and the total numbers of pluses and minuses are consistent with the signature of $\varepsilon_i P_{m_i}$. Since the diagram is defined up to the permutation of the rows of the same length, the unique type of degeneracy is determined by requiring the normalized ordering of the signs: among the rows of equal length, those with $\varepsilon_i=+1$ come first, followed by those with $\varepsilon_i=-1$.

An example of such a signed Young diagram is provided in Fig.~\ref{sydiag}. It corresponds to the type--$(5+,4+,4-,2-,-)$ EP. Using the theorem~\ref{theorem1}, the corresponding matrix and the pseudo-metric can be simultaneously brought into the forms
\begin{equation}
    H = T^{-1}J T,\quad J = J_5\oplus J_4\oplus J_4\oplus J_2\oplus J_1,
\end{equation}
\begin{multline}
    \eta = T^\dagger P T,\\ P = (+P_5)\oplus(+P_4)\oplus(-P_4)\oplus(-P_2)\oplus(-P_1).
\end{multline}
Counting the pluses and minuses in a row of the signed diagram, we get the numbers of positive and negative eigenvalues of the restriction $\varepsilon_i P_{m_i}$ of the pseudo-metric to the subspace of the corresponding Jordan block. Conversely, counting the total numbers of pluses and minuses in all of the rows, we get the number of positive and negative eigenvalues of the pseudo-metric. As such, for the case of the diagram on Fig.~\ref{sydiag}, the pseudo-metric must be $\eta_{8,8}$.

\begin{figure}
\ytableausetup{boxsize=1.5em}
\begin{ytableau}
+&-&+&-&+\\
-&+&-&+\\
+&-&+&-\\
+&-\\
-
\end{ytableau}
\caption{Example of a signed Young-diagram corresponding to the type--$(5+,4+,4-,2-,1-)$ EP.}
\label{sydiag}
\end{figure}

Let us now consider a general type-$(m_1\varepsilon_1, m_2\varepsilon_2,\dotsc,m_r\varepsilon_r)$ EP. The definition of the corresponding manifold of matrices has to be altered in comparison to the non-symmetric case of Section~\ref{GeometryN}, because the Jordan normal form itself generally does not belong to the manifold anymore.
Therefore, we choose some representative matrix $H_{m_1\varepsilon_1,\dotsc,m_r\varepsilon_p}$ hosting the given type of EP,
\begin{align}
   H_{m_1\varepsilon_1,\dotsc,m_r\varepsilon_r} &= T^{-1}J_{m_1,m_2,\dotsc,m_r}T,\label{eq:h-expand}\\
   \eta_{p,q} &= T^{\dagger}P_{m1\varepsilon_1,\dotsc,m_r\varepsilon_r}T,\label{eq:eta-expand}\\
   P_{m1\varepsilon_1,\dotsc,m_r\varepsilon_r} &=\bigoplus_{i=1}^{r} \varepsilon_i P_{m_i}.
\end{align}
Such a representative matrix can always be found: we merely need to solve Eq.~\eqref{eq:eta-expand} for $T$ and substitute the solution into Eq.~\eqref{eq:h-expand}.
The manifold of matrices hosting the given type of degeneracy is then obtained by acting on the representative matrix with similarity transformations preserving the indefinite scalar product associated with the pseudo-metric
\begin{equation}
    \BM_{m_1\varepsilon_1,\dotsc,m_r\varepsilon_r} = \{SH_{m_1\varepsilon_1,\dotsc,m_r\varepsilon_r}S^{-1}|S\in U(p,q)\}.
\end{equation}

In full analogy to the non-symmetric case, the hierarchy of EPs is defined through the partial dominance order that is induced on the set of degeneracy manifolds by the closure operation. For a type-$A$ and a type-$B$ EPs, $A\succ B\Leftrightarrow \BM_B\subset \overline{\BM_A}$. The condition $A\succ B$ implies that type-$B$ EP can be converted into type-$A$ EP by an infinitesimal perturbation. The dominance can be assessed by comparing the corresponding signed Young diagrams~\cite{Djokovic2006, Djokovic1982}:
Let $n^+(A)$ and $n^-(A)$ be the total numbers of pluses and minuses of the signed Young diagram $A$, respectively. In addition to that, let $A^{(k)}$ be the signed Young diagram obtained by removing the first $k$ columns of the diagram $A$. Then the condition of dominance is
\begin{equation}
    A\succ B\Leftrightarrow n^{\pm}(A^{(k)})\geqslant n^{\pm} (B^{(k)}),\, \, k=1,2\dotsc \, .
\end{equation}

If we denote by $n^\pm_l(A)$ the number of pluses (minuses) in the $l$-th column of diagram $A$, then
\begin{align}
    n^+(A^{(k)}) &= p-\sum_{l=1}^k n^+_l(A), \label{ipm1}\\
    n^-(A^{(k)}) &= q-\sum_{l=1}^k n^-_l(A), 
    \label{ipm2}
\end{align}
where $p=i^+(A)$ and $q=i^-(A)$ are determined by the pseudo-metric. Using Eqs.~\eqref{ipm1} and \eqref{ipm2}, we can also bring the condition of the dominance into a form convenient for calculations:
\begin{equation}
    A\succ B \, \Leftrightarrow  \, \sum_{l=1}^k n^{\pm}_l(A)\leqslant \sum_{l=1}^k n^{\pm}_l(B),\, \,  k=1,2\dotsc\,.
\end{equation}

As an example, we show the hierarchies of degeneracies in the cases of $\eta_{2,2}$ and $\eta_{3,1}$ pseudo-metrics in Figs.~\ref{eph422} and~\ref{eph431}, respectively. It is instructive to compare the Figs.~\ref{eph422} and~\ref{eph431} with the hierarchy Fig.~\ref{eph4} for algebraic multiplicity $4$ in the absence of symmetries. Ignoring the signs, the $\eta_{2,2}$-hierarchy contains the same diagrams as the hierarchy in Fig.~\ref{eph4}. For the $\eta_{3,1}$-hierarchy, we expect naturally that the $4$-th order non-derogatory EP is not possible, which is confirmed in Fig.~\ref{eph431}. Additionally, we observe that the diagram with two rows of size 2 is also absent, which lies in the middle of the hierarchy in Fig.~\ref{eph4}.

In conclusion of this section, we would like to mention that the \href{https://github.com/say-yas/Hierarchy_NH_degeneracies}{Mathematica notebook} and the \href{https://github.com/Gregstrq/DominanceHierarchy.jl}{Julia package} that we have developed are also capable of handling cases with pseudo-Hermitian symmetry.

\ytableausetup{boxsize=0.8em}
\begin{figure*}
\begin{tikzcd}[ampersand replacement=\&]
    \begin{ytableau}
        + & -& +& - \\
    \end{ytableau}
    \arrow[dr]\arrow[drrr]\arrow[dd]
    \& \& \& \&
    \begin{ytableau}
        - & +& -& + \\
    \end{ytableau}
    \arrow[dd]\arrow[dlll]\arrow[dl]
    \\
    \&
    \begin{ytableau}
        - & +& - \\
        +  \\
    \end{ytableau}
    \arrow[ddr]\arrow[ddrrr, bend left=10]\arrow[dr]
    \& \&
    \begin{ytableau}
        + & -& + \\
        -  \\
    \end{ytableau}
    \arrow[ddl]\arrow[ddr]\arrow[dl]
    \&
    \\
    \begin{ytableau}
        + & - \\
        + & - \\
    \end{ytableau}
    \arrow[drr]
    \& \&
    \begin{ytableau}
        - & + \\
        + & - \\
    \end{ytableau}
    \arrow[d]\arrow[drr]
    \& \&
    \begin{ytableau}
        - & + \\
        - & + \\
    \end{ytableau}
    \arrow[d]
    \\
    \& \&
    \begin{ytableau}
        + & - \\
        +  \\
        -  \\
    \end{ytableau}
    \arrow[dr]
    \& \&
    \begin{ytableau}
        - & + \\
        +  \\
        -  \\
    \end{ytableau}
    \arrow[dl]
    \\
    \& \& \&
    \begin{ytableau}
        +  \\
        +  \\
        -  \\
        -  \\
    \end{ytableau}    
    \&
    \\
\end{tikzcd}

\caption{Hierarchy of degeneracies with algebraic multiplicity $n=4$ corresponding to the pseudo-Hermitian symmetry with pseudo-metric $\eta_{2,2}$.}
\label{eph422}
\end{figure*}

\begin{figure}
\begin{tikzcd}[ampersand replacement=\&]
    \&
    \begin{ytableau}
        + & -& + \\
        +  \\
    \end{ytableau}
    \arrow[dl]\arrow[dr]
    \&
    \\
    \begin{ytableau}
        + & - \\
        +  \\
        +  \\
    \end{ytableau}
    \arrow[dr]
    \& \&
    \begin{ytableau}
        - & + \\
        +  \\
        +  \\
    \end{ytableau}
    \arrow[dl]
    \\
    \&
    \begin{ytableau}
        +  \\
        +  \\
        +  \\
        -  \\
    \end{ytableau}
    \&
    \\
\end{tikzcd}

\caption{Hierarchy of degeneracies with algebraic multiplicity $n=4$ corresponding to the pseudo-Hermitian symmetry with pseudo-metric $\eta_{3,1}$.}
\label{eph431}
\end{figure}

\section{EP conversion in eigenspectrum of Liouvillian Superoperator.}
\label{sec:applications}

Under general conditions, the density matrix of a system in contact with a dissipative environment satisfies the Lindblad master equation
\begin{multline}
    \frac{d}{dt}\hat \rho = \BL\qty[\hat \rho] = \\-i\qty[\hat H,\hat\rho]+\sum_{k}\Gamma_k\qty[\hat L_k \hat \rho \hat L_k^\dagger - \frac{1}{2}\{\hat L^\dagger_k\hat L_k,\hat\rho\}].\label{lindblad}
\end{multline}
Here, $\hat \rho$ and $\hat H$ are the density matrix and the Hamiltonian of the system, respectively, while $\hat L_k$ is a set of jump operators describing the influence of the environment on the system. We can recast the Lindblad equation as a matrix equation by vectorizing the density operator
\begin{equation}
    \hat \rho = \sum_{m,n}\rightarrow \bar \rho = \sum_{m,n} |m\rangle\otimes|n*\rangle.
\end{equation}
Here, we have used the fact that an operator acting on the Hilbert space $\BH$ can be identified with a vector from a tensor product space $\BH\otimes\BH^*$.

In the vectorized representation, the Lindblad equation casts
\begin{align}
    &\frac{d}{dt}\bar\rho =\BL\bar \rho,\\
    &\BL = \left[(-i\hat H_\mathrm{nh})\otimes\mathbb1 + \mathbb1\otimes(i\hat H_\mathrm{nh}^*)\right] + \sum_k \Gamma_k \hat L_k\otimes \hat L_k^*,\label{eq:lindblad-matrix}
\end{align}
where 
\begin{equation}
    \hat H_\mathrm{nh} = \hat H - \frac{1}{2}\sum_k \Gamma_k \hat L_k^\dagger\hat L_k.
\end{equation}
The first term in Eq.~\eqref{eq:lindblad-matrix} describes coherent dynamics with a non-Hermitian Hamiltonian $\hat H_\mathrm{nh}$, while the second one is called the quantum jump term and can be attributed to the sudden changes in the state of the system.
The matrix $\BL$ is referred to in the literature as a Liouvillian superoperator.

Let us focus on the part of the Liouvillian without quantum jumps, which reads
\begin{equation}
    \BL^\prime = (-i\hat H_\mathrm{nh})\otimes\mathbb1 + \mathbb1\otimes(i\hat H_\mathrm{nh}^*).\label{eq:no-jump-L}
\end{equation}
One can show that $\BL^\prime$ naturally exhibits a derogatory EP if the underlying non-Hermitian Hamiltonian $\hat H_\mathrm{nh}$ is itself tuned to one~\cite{shiralieva2025,trampus1966}.  For example, if $\hat H_\mathrm{nh}$ is tuned to an EP of maximal order $N$
\begin{equation}
    \hat H_\mathrm{nh}\sim J_N(\varepsilon),
\end{equation}
with $\varepsilon$ being the eigenvalue of $\hat H_\mathrm{nh}$. Subsequently, the no-jump Liouvillian is similar to
\begin{equation}
    \BL^\prime \sim J_{2N-1}(\lambda_\mathrm{EP})\oplus J_{2N-3}(\lambda_\mathrm{EP})\oplus\dots\oplus J_1(\lambda_\mathrm{EP}),
\end{equation}
where $\lambda_\mathrm{EP} = i(\varepsilon^*-\varepsilon)=2\im \varepsilon$.
Here $A\sim B$ means $A = S\,B\,S^{-1}$ for some invertible matrix $S$.

Taking into account the quantum jump term perturbs the derogatory multiblock EP. Generally, such a perturbation would split the eigenvalues. However, as we discussed in previous sections, it can also change the type of EP and merge some blocks. The hierarchy analysis introduced in the previous sections enables us to argue that certain EP conversions are theoretically possible without explicitly determining the form of the perturbation.

The Liouvillian in Eq.~\eqref{eq:lindblad-matrix} satisfies the generalized $\BP\BT$-symmetry~\cite{Sa_2023}, where $\check\BT$ is the complex conjugation operator and $\check\BP$ is the operator that swaps two terms in the kronecker products. We will additionally require that the Hamiltonian and the jump operators are symmetric (but possibly complex) matrices, so that the Liouvillian becomes $\BP$-pseudo-Hermitian, and subsequently, we can use the results of Section~\ref{GeometryPSH}.

As pointed out in the introduction, under generic perturbations, a derogatory EP behaves like a collection of non-derogatory EPs with the order determined by the sizes of the Jordan blocks. If we find a way to merge the blocks of the Linouvillian EP, we can effectively engineer a higher-order EP.

\subsubsection{Effective dissipative qubit}
As a concrete example, we will consider a three-level system with Hamiltonian
\begin{equation}
    \hat H = \sum_{l=1}^3 \varepsilon_l \ket{l}\bra{l} + \sum_{l<k}^{3} t_{lk}(\ket{l}\bra{k}+\ket{k}\bra{l}),
\end{equation}
subject to the dissipation affecting different groups of levels:
\begin{align}
    \BL=&\qty(-i\hat H)\otimes\mathbb1+\mathbb1\otimes\qty(i\hat H) \nonumber\\
    &+ \sum_{k=2}^3\gamma_{k} D\qty[\ket{1}\bra{k}] + \sum_{j}\Gamma_j \hat D\qty[L_j],\\
    D\qty[\hat L] =& \hat L\otimes\hat L^* - \frac{\hat L^\dagger\hat L\otimes\mathbb1 + \mathbb1\otimes\hat L^T\hat L^*}{2}.
\end{align}
Here, $\varepsilon_l$ sets the frequency detuning of the level $l$ and we consider  the coupling constants $t_{lk}$ to be real.
We assume that the quantum jump terms can be split into two groups: The terms of the first group describe dissipation from the excited levels into the ground state $| 1 \rangle$, and the corresponding jump rates are $\gamma_k$. The terms of the second group affect only the two excited states, and the corresponding jump rates are $\Gamma_j$.

We assume that there is no coherent coupling to the ground state $t_{1k}\equiv 0$. In this case, we can project out the ground state exactly
\begin{equation}
    \bar \rho_\mathrm{eff} = \check P \bar \rho,\quad \check P = \hat P\otimes\hat P^*,\quad \hat P = \sum_{l\neq 1} \ket{l}\bra{l}.
\end{equation}
Experimentally, this can be achieved by the post-selection techniques~\cite{naghiloo2019, chen_quantum_2021}.

The projected Liouvillian describes the dynamics of the two excited levels that constitute an effective non-Hermitian qubit:
\begin{align}
    \frac{d}{dt}\bar \rho_\mathrm{eff} &= \BL_\mathrm{eff}\bar\rho,\\
    \BL_\mathrm{eff} &= \check P\BL\check P=\qty[(-i\hat H_\mathrm{eff})\otimes\mathbb1+\mathbb1\otimes(i\hat H_\mathrm{eff}^*)]\nonumber\\ &\hspace{1.6cm}+ \sum_{j}\Gamma_{j}D\qty[\hat L_j],\label{eq:lindblad-proj}\\
    \hat H_\mathrm{eff} &= \begin{pmatrix}
        \varepsilon_2-\frac{i\gamma_2}{2} & t\\
        t & \varepsilon_3-\frac{i\gamma_3}{2}
    \end{pmatrix}.
\end{align}
Here, we did not distinguish between projected and unprojected $\hat L_j$, since by assumption they affect only excited states, $\check P D[\hat L_j]\check P = D[\hat L_j]$.

The step with the projection might seem superfluous; however, it allows us to decouple the effective non-Hermitian Hamiltonian $\hat H_{\rm nh}$ from the quantum jump term in Eq.~\eqref{eq:lindblad-matrix} to some extent. As such, even if we neglect the dissipative terms $\hat L_j$, the dynamics of the effective qubit is still described by a non-Hermitian Hamiltonian $\hat H_{\rm eff}$.

We can tune $\hat H_\mathrm{eff}$ to a second-order EP, if we choose $\varepsilon_2=\varepsilon_3$ and $\gamma_2-\gamma_3 = \pm 4t$. In this case, the no-jump part of $\BL_\mathrm{eff}$ casts
\begin{align}
    \BL^\prime_\mathrm{eff} &= \qty[(-i\hat H_\mathrm{eff})\otimes\mathbb1+\mathbb1\otimes(i\hat H_\mathrm{eff}^*)], \nonumber \\
    &= -\frac{\gamma_2+\gamma_3}{2}I_4 + \begin{pmatrix}
        \mp 2t & it & -it & 0\\
        it & 0 & 0 & -it\\
        -it & 0 & 0 & it\\
        0 & -it & it & \pm 2t
    \end{pmatrix}.
\end{align}

The parity operator $\check\BP$ acts on the vectorized states as
\begin{equation}
    \check\BP |m\rangle\otimes|n^*\rangle = |n\rangle\otimes|m^*\rangle.
\end{equation}
In the same basis as $\BL^\prime_\mathrm{eff}$, we can write the explicit matrix form of $\check\BP$ as
\begin{equation}
    \check\BP=\begin{pmatrix}
        1 & 0 & 0 & 0\\
        0 & 0 & 1 & 0\\
        0 & 1 & 0 & 0\\
        0 & 0 & 0 & 1
    \end{pmatrix}.
\end{equation}
One can also check that $\BL_\mathrm{eff}^\prime$ is indeed pseudo-Hermitian with respect to $\check\BP$.

Using this information, we can determine the type of EP exhibited by $\BL^\prime_\mathrm{eff}$, including the signs. Theoretically, one should construct the basis from Theorem~\ref{theo} to do this. However, in our case, we can adopt a simpler approach.

First, we find the signature of the pseudo-metric using
\begin{equation}
    p-q = \tr\qty[\check \BP] = 2,
\end{equation}
which results in
\begin{equation}
    \check \BP\sim \eta_{3,1},
\end{equation}
with the corresponding hierarchy shown in Fig.~\ref{eph431}.
Secondly, we already know that $\BL^\prime_\mathrm{eff}$ exhibits a derogatory EP with block sizes $3$ and $1$. Looking at Fig.~\ref{eph431}, we identify the EP as type-$(3+,1+)$. Notably, there is no EP with a single block of size $4$ in the hierarchy. Therefore, we conclude that regardless of the type of dissipative terms $D\qty[\hat L_j]$ we include in $\BL^\prime_\mathrm{eff}$, as long as the $\BP$-pseudo-Hermitian symmetry is maintained, it will be impossible to merge the blocks.

\subsubsection{Effective non-Hermitian qutrit.}

In an analogous manner, we can consider a four-level system with dissipation into the ground state, which we project out to get an effective non-Hermitian qutrit.

The projected Liouvillian has the same form~\eqref{eq:lindblad-proj} with a different $\hat H_\mathrm{eff}$ which reads
\begin{equation}
    \hat H_\mathrm{eff} = \begin{pmatrix}
        \varepsilon_2-\frac{i\gamma_2}{2} & t_{23} & t_{24}\\
        t_{23} & \varepsilon_3-\frac{i\gamma_3}{2} & t_{34}\\
        t_{24} & t_{34} & \varepsilon_4-\frac{i\gamma_4}{2}
    \end{pmatrix}.
\end{equation}
The jump operators $\hat L_j$ are assumed to act only on the levels of the effective qutrit.

The non-Hermitian Hamitlonian $\hat H_\mathrm{eff}$ is tuned to a non-derogatory EP3 when $t_{24}=0$, $t_{23}=t_{34}=t$, $\varepsilon_{2,3,4}\equiv \varepsilon$, $2\gamma_3 = \gamma_2+\gamma_4$ and $\gamma_2-\gamma_4=\pm4\sqrt{2}t$~\cite{shiralieva2025}. In this case, the no-jump part of $\BL_\mathrm{eff}$ becomes
\begin{widetext}
\begin{equation}
    \BL_\mathrm{eff}^\prime = \qty[(-i\hat H_\mathrm{eff})\otimes\mathbb1+\mathbb1\otimes(i\hat H_\mathrm{eff}^*)] = -\gamma_i \mathbb1_9 
    +t
\begin{pmatrix}
 \mp2 \sqrt{2} & i & 0 & -i & 0 & 0 & 0 & 0 & 0 \\
 i & \mp\sqrt{2} & i & 0 & -i & 0 & 0 & 0 & 0 \\
 0 & i & 0 & 0 & 0 & -i & 0 & 0 & 0 \\
 -i & 0 & 0 & \mp\sqrt{2} & i & 0 & -i & 0 & 0 \\
 0 & -i & 0 & i & 0 & i & 0 & -i & 0 \\
 0 & 0 & -i & 0 & i & \pm\sqrt{2} & 0 & 0 & -i \\
 0 & 0 & 0 & -i & 0 & 0 & 0 & i & 0 \\
 0 & 0 & 0 & 0 & -i & 0 & i & \pm\sqrt{2} & i \\
 0 & 0 & 0 & 0 & 0 & -i & 0 & i & \pm2 \sqrt{2}
\end{pmatrix}.\label{eq:eff-qutrit-ep}
\end{equation}
\end{widetext}
It exhibits a derogatory EP characterized by the Jordan normal form with block sizes $(5,3,1)$:
\begin{multline}
    \BL_\mathrm{eff}^\prime = \qty[(-i\hat H_\mathrm{eff})\otimes\mathbb1+\mathbb1\otimes(i\hat H_\mathrm{eff}^*)]\sim\\
    \sim
    J_5(\lambda_\mathrm{EP})\oplus J_3(\lambda_\mathrm{EP})\oplus J_1(\lambda_\mathrm{EP}),
\end{multline}
where $\lambda_\mathrm{EP}=-\gamma_3$.

The basis used in  Eq.~\eqref{eq:eff-qutrit-ep} is enumerated as
\begin{multline}
    \ket{2}\otimes\ket{2^*}, \ket{2}\otimes\ket{3^*}, \ket{2}\otimes\ket{4^*}, \ket{3}\otimes\ket{2^*}, \ket{3}\otimes\ket{3^*},\\ \ket{3}\otimes\ket{4^*},\ket{4}\otimes\ket{2^*},\ket{4}\otimes\ket{3^*},\ket{4}\otimes\ket{4^*}.
\end{multline}
The parity operator $\check \BP$ swaps second and fourth, third and seventh, sixth and eighth basis vectors, while doing nothing with the rest of the basis vectors. This information allows us to write the explicit matrix form of $\check\BP$
\begin{equation}
    \check\BP = \begin{pmatrix}
        1 & 0 & 0 & 0 & 0 & 0 & 0 & 0 & 0 \\
        0 & 0 & 0 & 1 & 0 & 0 & 0 & 0 & 0 \\
        0 & 0 & 0 & 0 & 0 & 0 & 1 & 0 & 0 \\
        0 & 1 & 0 & 0 & 0 & 0 & 0 & 0 & 0 \\
        0 & 0 & 0 & 0 & 1 & 0 & 0 & 0 & 0 \\
        0 & 0 & 0 & 0 & 0 & 0 & 0 & 1 & 0 \\
        0 & 0 & 1 & 0 & 0 & 0 & 0 & 0 & 0 \\
        0 & 0 & 0 & 0 & 0 & 1 & 0 & 0 & 0 \\
        0 & 0 & 0 & 0 & 0 & 0 & 0 & 0 & 1
    \end{pmatrix}.
\end{equation}
One can easily check that $\BL_\mathrm{eff}^\prime$ is indeed pseudo-Hermitian with respect to $\check\BP$.

\begin{figure*}[p]
\includegraphics[angle=90,width=0.55\textwidth]{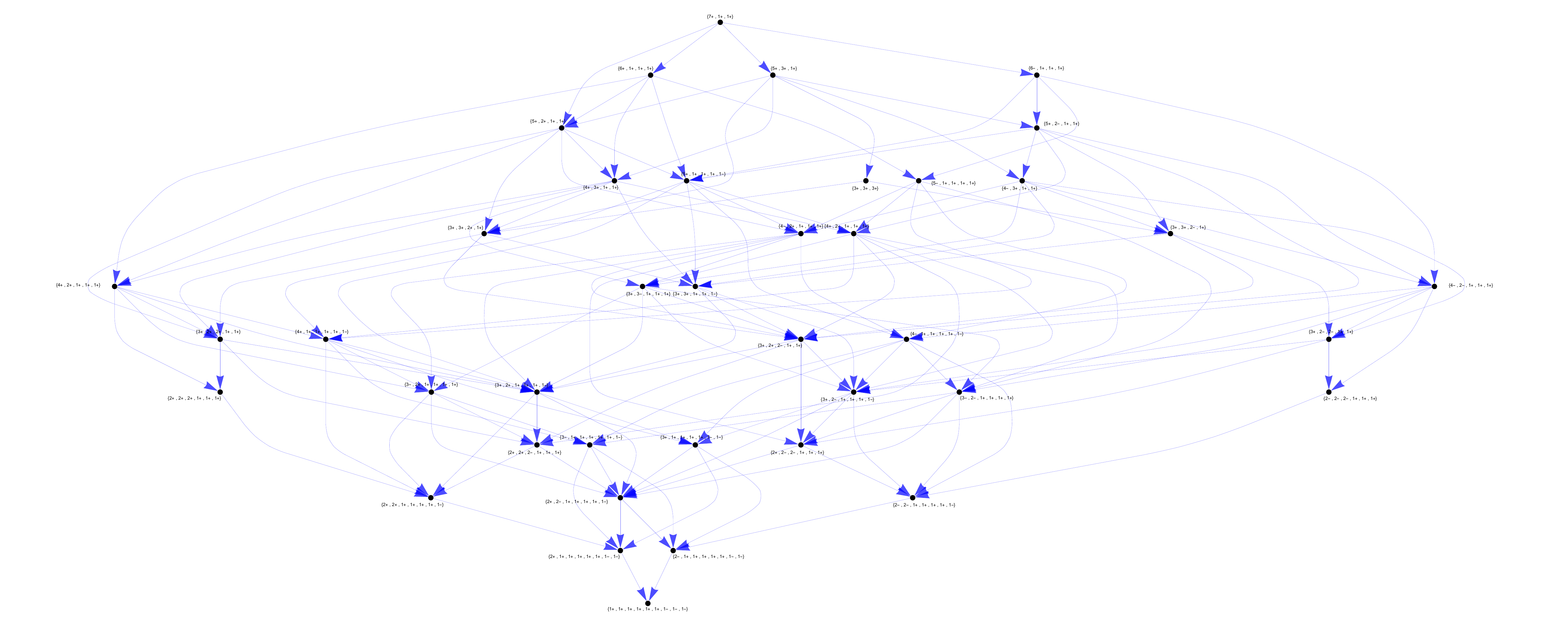}
\caption{Hierarchy of degeneracies with algebraic multiplicity $n=9$ corresponding to the pseudo-Hermitian symmetry with pseudo-metric $\eta_{6,3}$.}
\label{eph963}
\end{figure*}

Using the explicit form of $\check\BP$, we can determine the signature of the pseudo-metric such that
\begin{equation}
    p-q = \tr\qty[\check\BP] = 3,
\end{equation}
so the canonical form of the pseudo-metric is $\eta_{6,3}$.
We present the corresponding hierarchy in Fig.~\ref{eph963}. There is only one diagram with Jordan block sizes $(5,3,1)$, which allows us to identify the type of EP as $(5+,3+,1+)$ (the middle diagram in the second row of the top). There is also a single diagram that is higher in the hierarchy: $(7+,1+,1+)$. Therefore, by incorporating dissipation terms $D\qty[\hat L_j]$ to $\BL_\mathrm{eff}^\prime$, it should theoretically be possible to partially merge the blocks and get a derogatory EP with the largest Jordan block of size $7$.

It is important to note that we can not claim that a combination of the dissipation terms that renders this type of EP actually exists: the terms $D\qty[\hat L_j]$ have restricted form and do not constitute a general perturbation of $\BL_\mathrm{eff}^\prime$ consistent with the $\check P$-pseudo-Hermitian symmetry. 

\section{Discussion and Conclusions}
\label{sec:conclusion} 

In this work, we have demonstrated a unique feature universal to the derogatory exceptional points (EPs), which are characterized by multiple Jordan blocks corresponding to the same eigenvalue. In many instances, infinitesimal perturbations can convert these EPs into EPs of other types without altering the order of the degeneracy. In these regards, a non-defective degeneracy ($n$-bolical point) can be viewed as an extreme case of a derogatory EP, as it can be converted to any type of EP through an appropriate perturbation.

The sensitivity of a derogatory EP to changes in parameters, which is an essential property for the practical application of EP-based devices, depends on the size of the largest Jordan block, which can be increased through the EP conversion. On the other hand, engineering specific EP conversion could be complicated. Therefore, it is crucial to understand which transformations are possible without determining the specific form of the perturbation. The hierarchy of EPs serves as a valuable tool for determining which conversions are feasible. In our work, we have demonstrated how such hierarchies can be constructed in cases where no symmetry exists and in situations involving pseudo-Hermitian symmetry. Moreover, we have also developed a \href{https://github.com/say-yas/Hierarchy_NH_degeneracies}{Mathematica notebook} and a \href{https://github.com/Gregstrq/DominanceHierarchy.jl}{Julia package} that automate this type of analysis.

An interesting platform for the application of our findings is the eigenspectra of Liouvillean superoperators.
As it was shown in Ref.~\cite{shiralieva2025}, Liouvillian superoperator naturally hosts a derogatory EP, if the underlying effective non-Hermitian Hamiltonian is itself tuned to an EP and the quantum jump terms are neglected. This poses the following interesting question: can we increase the size of the largest Jordan block by converting the Liouvillian EP through engineering the quantum jump terms?
The dimensionality of the Liouvillian is the square of the dimensionality of the effective non-Hermitian Hamiltonian, so potentially the size of the largest Jordan block can be made quite large.

In order to apply the hierarchy analysis to the most general Liouvillian superoperators, we need to figure out how to extend the results in the case of the pseudo-Hermitian symmetry to the case of $\BP\BT$-symmetry. The main complication here is the following. Any $\BP\BT$-symmetric matrix is pseudo-Hermitian in the sense that we can find a suitable pseudo-metric operator satisfying condition~\eqref{pH-def}. However, its choice is not unique~\cite{bian2020conserved,agarwal2022conserved} and depends on the parameters of the system~\cite{mostafazadeh2002pseudo,mostafazadeh2,mostafazadeh3,mostafazadeh2010, Ashida2020non,ZhangQinXiao-20}. In order to study small parameter perturbations, one needs to fix the choice of pseudo-metric that would be parameter-independent (which is equivalent to the case of the pseudo-Hermitian symmetry) or at least change with parameters in a continuous manner. Whether it is possible in all cases is an open question. We hope to address this in future work.

Another general way to realize a derogatory EP is by looking at a composite system~\cite{Wiersig2025}. The idea here is to have two non-interacting copies with the same Hamiltonian $\hat H_\mathrm{EP}$ tuned to an EP. This casts the total Hamiltonian of the system as
\begin{equation}
    \hat H_\mathrm{tot} = \hat H\otimes\mathbb 1+\mathbb{1}\otimes \hat H.
\end{equation}
Clearly, $\hat H_\mathrm{tot} $ has a similar structure to the Liouvillian superoperator~\eqref{eq:no-jump-L} without quantum jumps. Thereby, following our discussion in the main text, it may also host a derogatory EP.
When we introduce interaction or coupling terms between the two copies, the EP blocks may, in principle, merge together resulting in the EP conversion.

\section{Acknowledgement}
G.~S. is greatly indebted to I. Danilenko for the enlightening discussions, and would like to thank A. Shiralieva.
S.~S. acknowledges fruitful discussions and insightful communications with D.~Bahns, M.~Brandt, F.~Mohammadi and B.~Strumfels.
G.~S. is supported by DFG-SFB 1170 (Project-ID: 258499086) and EXC2147 ctd.qmat (Project-ID: 390858490).

 \bibliography{bibfile}

\begin{thebibliography}{79}%
\makeatletter
\providecommand \@ifxundefined [1]{%
 \@ifx{#1\undefined}
}%
\providecommand \@ifnum [1]{%
 \ifnum #1\expandafter \@firstoftwo
 \else \expandafter \@secondoftwo
 \fi
}%
\providecommand \@ifx [1]{%
 \ifx #1\expandafter \@firstoftwo
 \else \expandafter \@secondoftwo
 \fi
}%
\providecommand \natexlab [1]{#1}%
\providecommand \enquote  [1]{``#1''}%
\providecommand \bibnamefont  [1]{#1}%
\providecommand \bibfnamefont [1]{#1}%
\providecommand \citenamefont [1]{#1}%
\providecommand \href@noop [0]{\@secondoftwo}%
\providecommand \href [0]{\begingroup \@sanitize@url \@href}%
\providecommand \@href[1]{\@@startlink{#1}\@@href}%
\providecommand \@@href[1]{\endgroup#1\@@endlink}%
\providecommand \@sanitize@url [0]{\catcode `\\12\catcode `\$12\catcode
  `\&12\catcode `\#12\catcode `\^12\catcode `\_12\catcode `\%12\relax}%
\providecommand \@@startlink[1]{}%
\providecommand \@@endlink[0]{}%
\providecommand \url  [0]{\begingroup\@sanitize@url \@url }%
\providecommand \@url [1]{\endgroup\@href {#1}{\urlprefix }}%
\providecommand \urlprefix  [0]{URL }%
\providecommand \Eprint [0]{\href }%
\providecommand \doibase [0]{http://dx.doi.org/}%
\providecommand \selectlanguage [0]{\@gobble}%
\providecommand \bibinfo  [0]{\@secondoftwo}%
\providecommand \bibfield  [0]{\@secondoftwo}%
\providecommand \translation [1]{[#1]}%
\providecommand \BibitemOpen [0]{}%
\providecommand \bibitemStop [0]{}%
\providecommand \bibitemNoStop [0]{.\EOS\space}%
\providecommand \EOS [0]{\spacefactor3000\relax}%
\providecommand \BibitemShut  [1]{\csname bibitem#1\endcsname}%
\let\auto@bib@innerbib\@empty
\bibitem [{\citenamefont {Ashida}\ \emph {et~al.}(2020)\citenamefont {Ashida},
  \citenamefont {Gong},\ and\ \citenamefont {Ueda}}]{Ashida2020non}%
  \BibitemOpen
  \bibfield  {author} {\bibinfo {author} {\bibfnamefont {Yuto}\ \bibnamefont
  {Ashida}}, \bibinfo {author} {\bibfnamefont {Zongping}\ \bibnamefont {Gong}},
  \ and\ \bibinfo {author} {\bibfnamefont {Masahito}\ \bibnamefont {Ueda}},\
  }\bibfield  {title} {\enquote {\bibinfo {title} {Non-hermitian physics},}\
  }\href {https://doi.org/10.1080/00018732.2021.1876991} {\bibfield  {journal}
  {\bibinfo  {journal} {Advances in Physics}\ }\textbf {\bibinfo {volume}
  {69}},\ \bibinfo {pages} {249--435} (\bibinfo {year} {2020})}\BibitemShut
  {NoStop}%
\bibitem [{\citenamefont {Wang}\ \emph {et~al.}(2021)\citenamefont {Wang},
  \citenamefont {Zhang}, \citenamefont {Hua}, \citenamefont {Lei},
  \citenamefont {Lu},\ and\ \citenamefont {Chen}}]{Wang_2021}%
  \BibitemOpen
  \bibfield  {author} {\bibinfo {author} {\bibfnamefont {Hongfei}\ \bibnamefont
  {Wang}}, \bibinfo {author} {\bibfnamefont {Xiujuan}\ \bibnamefont {Zhang}},
  \bibinfo {author} {\bibfnamefont {Jinguo}\ \bibnamefont {Hua}}, \bibinfo
  {author} {\bibfnamefont {Dangyuan}\ \bibnamefont {Lei}}, \bibinfo {author}
  {\bibfnamefont {Minghui}\ \bibnamefont {Lu}}, \ and\ \bibinfo {author}
  {\bibfnamefont {Yanfeng}\ \bibnamefont {Chen}},\ }\bibfield  {title}
  {\enquote {\bibinfo {title} {Topological physics of non-hermitian optics and
  photonics: a review},}\ }\href {\doibase 10.1088/2040-8986/ac2e15} {\bibfield
   {journal} {\bibinfo  {journal} {Journal of Optics}\ }\textbf {\bibinfo
  {volume} {23}},\ \bibinfo {pages} {123001} (\bibinfo {year}
  {2021})}\BibitemShut {NoStop}%
\bibitem [{\citenamefont {Miri}\ and\ \citenamefont
  {Alu}(2019)}]{Miri2019exceptional}%
  \BibitemOpen
  \bibfield  {author} {\bibinfo {author} {\bibfnamefont {Mohammad-Ali}\
  \bibnamefont {Miri}}\ and\ \bibinfo {author} {\bibfnamefont {Andrea}\
  \bibnamefont {Alu}},\ }\bibfield  {title} {\enquote {\bibinfo {title}
  {Exceptional points in optics and photonics},}\ }\href {\doibase
  10.1126/science.aar7709} {\bibfield  {journal} {\bibinfo  {journal}
  {Science}\ }\textbf {\bibinfo {volume} {363}},\ \bibinfo {pages} {eaar7709}
  (\bibinfo {year} {2019})}\BibitemShut {NoStop}%
\bibitem [{\citenamefont {Arkhipov}\ \emph {et~al.}(2024)\citenamefont
  {Arkhipov}, \citenamefont {Minganti}, \citenamefont {Miranowicz},
  \citenamefont {\"Ozdemir},\ and\ \citenamefont {Nori}}]{Arkhipov2024}%
  \BibitemOpen
  \bibfield  {author} {\bibinfo {author} {\bibfnamefont {Ievgen~I.}\
  \bibnamefont {Arkhipov}}, \bibinfo {author} {\bibfnamefont {Fabrizio}\
  \bibnamefont {Minganti}}, \bibinfo {author} {\bibfnamefont {Adam}\
  \bibnamefont {Miranowicz}}, \bibinfo {author} {\bibfnamefont {\ifmmode
  \mbox{\c{S}}\else \c{S}\fi{}ahin~K.}\ \bibnamefont {\"Ozdemir}}, \ and\
  \bibinfo {author} {\bibfnamefont {Franco}\ \bibnamefont {Nori}},\ }\bibfield
  {title} {\enquote {\bibinfo {title} {Restoring adiabatic state transfer in
  time-modulated non-hermitian systems},}\ }\href {\doibase
  10.1103/PhysRevLett.133.113802} {\bibfield  {journal} {\bibinfo  {journal}
  {Phys. Rev. Lett.}\ }\textbf {\bibinfo {volume} {133}},\ \bibinfo {pages}
  {113802} (\bibinfo {year} {2024})}\BibitemShut {NoStop}%
\bibitem [{\citenamefont {Wu}\ \emph {et~al.}(2025)\citenamefont {Wu},
  \citenamefont {Zhao}, \citenamefont {Zhou}, \citenamefont {Ye}, \citenamefont
  {Fang}, \citenamefont {Zhou},\ and\ \citenamefont {Yang}}]{Wu-25}%
  \BibitemOpen
  \bibfield  {author} {\bibinfo {author} {\bibfnamefont {Qi-Cheng}\
  \bibnamefont {Wu}}, \bibinfo {author} {\bibfnamefont {Jun-Long}\ \bibnamefont
  {Zhao}}, \bibinfo {author} {\bibfnamefont {Yan-Hui}\ \bibnamefont {Zhou}},
  \bibinfo {author} {\bibfnamefont {Biao-Liang}\ \bibnamefont {Ye}}, \bibinfo
  {author} {\bibfnamefont {Yu-Liang}\ \bibnamefont {Fang}}, \bibinfo {author}
  {\bibfnamefont {Zheng-Wei}\ \bibnamefont {Zhou}}, \ and\ \bibinfo {author}
  {\bibfnamefont {Chui-Ping}\ \bibnamefont {Yang}},\ }\bibfield  {title}
  {\enquote {\bibinfo {title} {Shortcuts to adiabatic state transfer in
  time-modulated two-level non-hermitian systems},}\ }\href {\doibase
  10.1103/PhysRevA.111.022410} {\bibfield  {journal} {\bibinfo  {journal}
  {Phys. Rev. A}\ }\textbf {\bibinfo {volume} {111}},\ \bibinfo {pages}
  {022410} (\bibinfo {year} {2025})}\BibitemShut {NoStop}%
\bibitem [{\citenamefont {Bai}\ \emph {et~al.}(2025)\citenamefont {Bai},
  \citenamefont {Lin}, \citenamefont {Liu}, \citenamefont {Li}, \citenamefont
  {Lyu},\ and\ \citenamefont {Xiao}}]{Bai-25}%
  \BibitemOpen
  \bibfield  {author} {\bibinfo {author} {\bibfnamefont {Kai}\ \bibnamefont
  {Bai}}, \bibinfo {author} {\bibfnamefont {Chen}\ \bibnamefont {Lin}},
  \bibinfo {author} {\bibfnamefont {Tao}\ \bibnamefont {Liu}}, \bibinfo
  {author} {\bibfnamefont {Jia-Zheng}\ \bibnamefont {Li}}, \bibinfo {author}
  {\bibfnamefont {Xin}\ \bibnamefont {Lyu}}, \ and\ \bibinfo {author}
  {\bibfnamefont {Meng}\ \bibnamefont {Xiao}},\ }\bibfield  {title} {\enquote
  {\bibinfo {title} {Nonlinear chiral-like state transfer realized with a
  minimal set of parameters},}\ }\href {\doibase 10.1038/s41467-025-61372-2}
  {\bibfield  {journal} {\bibinfo  {journal} {Nature Communications}\ }\textbf
  {\bibinfo {volume} {16}} (\bibinfo {year} {2025}),\
  10.1038/s41467-025-61372-2}\BibitemShut {NoStop}%
\bibitem [{\citenamefont {Laha}\ \emph {et~al.}(2025)\citenamefont {Laha},
  \citenamefont {Beniwal}, \citenamefont {Ghosh},\ and\ \citenamefont
  {Miranowicz}}]{Laha-25}%
  \BibitemOpen
  \bibfield  {author} {\bibinfo {author} {\bibfnamefont {Arnab}\ \bibnamefont
  {Laha}}, \bibinfo {author} {\bibfnamefont {Dinesh}\ \bibnamefont {Beniwal}},
  \bibinfo {author} {\bibfnamefont {Somnath}\ \bibnamefont {Ghosh}}, \ and\
  \bibinfo {author} {\bibfnamefont {Adam}\ \bibnamefont {Miranowicz}},\
  }\bibfield  {title} {\enquote {\bibinfo {title} {Programmable state switching
  based on higher-order exceptional points in anti-parity-time symmetric
  microcavity systems},}\ }\href {\doibase 10.1038/s41598-025-13797-4}
  {\bibfield  {journal} {\bibinfo  {journal} {Scientific Reports}\ }\textbf
  {\bibinfo {volume} {15}} (\bibinfo {year} {2025}),\
  10.1038/s41598-025-13797-4}\BibitemShut {NoStop}%
\bibitem [{\citenamefont {Qu}\ \emph {et~al.}(2026)\citenamefont {Qu},
  \citenamefont {Arkhipov}, \citenamefont {Gao}, \citenamefont {Wang},
  \citenamefont {Xiao}, \citenamefont {Nori},\ and\ \citenamefont
  {Xue}}]{Qu-26}%
  \BibitemOpen
  \bibfield  {author} {\bibinfo {author} {\bibfnamefont {Dengke}\ \bibnamefont
  {Qu}}, \bibinfo {author} {\bibfnamefont {Ievgen~I.}\ \bibnamefont
  {Arkhipov}}, \bibinfo {author} {\bibfnamefont {Huixia}\ \bibnamefont {Gao}},
  \bibinfo {author} {\bibfnamefont {Kunkun}\ \bibnamefont {Wang}}, \bibinfo
  {author} {\bibfnamefont {Lei}\ \bibnamefont {Xiao}}, \bibinfo {author}
  {\bibfnamefont {Franco}\ \bibnamefont {Nori}}, \ and\ \bibinfo {author}
  {\bibfnamefont {Peng}\ \bibnamefont {Xue}},\ }\bibfield  {title} {\enquote
  {\bibinfo {title} {Selective chiral multistate switching via the dynamic
  interplay of diabolic and exceptional points},}\ }\href {\doibase
  10.1103/wvrz-432c} {\bibfield  {journal} {\bibinfo  {journal} {Phys. Rev.
  Lett.}\ }\textbf {\bibinfo {volume} {136}},\ \bibinfo {pages} {086603}
  (\bibinfo {year} {2026})}\BibitemShut {NoStop}%
\bibitem [{\citenamefont {Hodaei}\ \emph {et~al.}(2017)\citenamefont {Hodaei},
  \citenamefont {Hassan}, \citenamefont {Wittek}, \citenamefont
  {Garcia-Gracia}, \citenamefont {El-Ganainy}, \citenamefont
  {Christodoulides},\ and\ \citenamefont {Khajavikhan}}]{Khajavikhan-17}%
  \BibitemOpen
  \bibfield  {author} {\bibinfo {author} {\bibfnamefont {Hossein}\ \bibnamefont
  {Hodaei}}, \bibinfo {author} {\bibfnamefont {Absar~U.}\ \bibnamefont
  {Hassan}}, \bibinfo {author} {\bibfnamefont {Steffen}\ \bibnamefont
  {Wittek}}, \bibinfo {author} {\bibfnamefont {Hipolito}\ \bibnamefont
  {Garcia-Gracia}}, \bibinfo {author} {\bibfnamefont {Ramy}\ \bibnamefont
  {El-Ganainy}}, \bibinfo {author} {\bibfnamefont {Demetrios~N.}\ \bibnamefont
  {Christodoulides}}, \ and\ \bibinfo {author} {\bibfnamefont {Mercedeh}\
  \bibnamefont {Khajavikhan}},\ }\bibfield  {title} {\enquote {\bibinfo {title}
  {Enhanced sensitivity at higher-order exceptional points},}\ }\href {\doibase
  10.1038/nature23280} {\bibfield  {journal} {\bibinfo  {journal} {Nature}\
  }\textbf {\bibinfo {volume} {548}},\ \bibinfo {pages} {187--191} (\bibinfo
  {year} {2017})}\BibitemShut {NoStop}%
\bibitem [{\citenamefont {Lai}\ \emph {et~al.}(2019)\citenamefont {Lai},
  \citenamefont {Lu}, \citenamefont {Suh}, \citenamefont {Yuan},\ and\
  \citenamefont {Vahala}}]{lai2019observation}%
  \BibitemOpen
  \bibfield  {author} {\bibinfo {author} {\bibfnamefont {Yu-Hung}\ \bibnamefont
  {Lai}}, \bibinfo {author} {\bibfnamefont {Yu-Kun}\ \bibnamefont {Lu}},
  \bibinfo {author} {\bibfnamefont {Myoung-Gyun}\ \bibnamefont {Suh}}, \bibinfo
  {author} {\bibfnamefont {Zhiquan}\ \bibnamefont {Yuan}}, \ and\ \bibinfo
  {author} {\bibfnamefont {Kerry}\ \bibnamefont {Vahala}},\ }\bibfield  {title}
  {\enquote {\bibinfo {title} {Observation of the exceptional-point-enhanced
  sagnac effect},}\ }\href {https://doi.org/10.1038/s41586-019-1777-z}
  {\bibfield  {journal} {\bibinfo  {journal} {Nature}\ }\textbf {\bibinfo
  {volume} {576}},\ \bibinfo {pages} {65--69} (\bibinfo {year}
  {2019})}\BibitemShut {NoStop}%
\bibitem [{\citenamefont {Kononchuk}\ \emph {et~al.}(2022)\citenamefont
  {Kononchuk}, \citenamefont {Cai}, \citenamefont {Ellis}, \citenamefont
  {Thevamaran},\ and\ \citenamefont {Kottos}}]{kononchuk2022exceptional}%
  \BibitemOpen
  \bibfield  {author} {\bibinfo {author} {\bibfnamefont {Rodion}\ \bibnamefont
  {Kononchuk}}, \bibinfo {author} {\bibfnamefont {Jizhe}\ \bibnamefont {Cai}},
  \bibinfo {author} {\bibfnamefont {Fred}\ \bibnamefont {Ellis}}, \bibinfo
  {author} {\bibfnamefont {Ramathasan}\ \bibnamefont {Thevamaran}}, \ and\
  \bibinfo {author} {\bibfnamefont {Tsampikos}\ \bibnamefont {Kottos}},\
  }\bibfield  {title} {\enquote {\bibinfo {title} {Exceptional-point-based
  accelerometers with enhanced signal-to-noise ratio},}\ }\href
  {https://doi.org/10.1038/s41586-022-04904-w} {\bibfield  {journal} {\bibinfo
  {journal} {Nature}\ }\textbf {\bibinfo {volume} {607}},\ \bibinfo {pages}
  {697--702} (\bibinfo {year} {2022})}\BibitemShut {NoStop}%
\bibitem [{\citenamefont {Wang}\ \emph {et~al.}(2025)\citenamefont {Wang},
  \citenamefont {Liang}, \citenamefont {Hu}, \citenamefont {Zhou},
  \citenamefont {Liu}, \citenamefont {Hu}, \citenamefont {Wang},\ and\
  \citenamefont {Ye}}]{Wang-24}%
  \BibitemOpen
  \bibfield  {author} {\bibinfo {author} {\bibfnamefont {Zuoxian}\ \bibnamefont
  {Wang}}, \bibinfo {author} {\bibfnamefont {Zihua}\ \bibnamefont {Liang}},
  \bibinfo {author} {\bibfnamefont {Jinsheng}\ \bibnamefont {Hu}}, \bibinfo
  {author} {\bibfnamefont {Peng}\ \bibnamefont {Zhou}}, \bibinfo {author}
  {\bibfnamefont {Lu}~\bibnamefont {Liu}}, \bibinfo {author} {\bibfnamefont
  {Gen}\ \bibnamefont {Hu}}, \bibinfo {author} {\bibfnamefont {Weiyi}\
  \bibnamefont {Wang}}, \ and\ \bibinfo {author} {\bibfnamefont {Mao}\
  \bibnamefont {Ye}},\ }\bibfield  {title} {\enquote {\bibinfo {title} {Sensing
  applications of pt-symmetry in non-hermitian photonic systems},}\ }\href
  {\doibase https://doi.org/10.1002/qute.202400349} {\bibfield  {journal}
  {\bibinfo  {journal} {Advanced Quantum Technologies}\ }\textbf {\bibinfo
  {volume} {8}},\ \bibinfo {pages} {2400349} (\bibinfo {year}
  {2025})}\BibitemShut {NoStop}%
\bibitem [{\citenamefont {Parto}\ \emph {et~al.}(2025)\citenamefont {Parto},
  \citenamefont {Leefmans}, \citenamefont {Williams}, \citenamefont {Gray},\
  and\ \citenamefont {Marandi}}]{Parto2025enhanced}%
  \BibitemOpen
  \bibfield  {author} {\bibinfo {author} {\bibfnamefont {Midya}\ \bibnamefont
  {Parto}}, \bibinfo {author} {\bibfnamefont {Christian}\ \bibnamefont
  {Leefmans}}, \bibinfo {author} {\bibfnamefont {James}\ \bibnamefont
  {Williams}}, \bibinfo {author} {\bibfnamefont {Robert~M}\ \bibnamefont
  {Gray}}, \ and\ \bibinfo {author} {\bibfnamefont {Alireza}\ \bibnamefont
  {Marandi}},\ }\bibfield  {title} {\enquote {\bibinfo {title} {Enhanced
  sensitivity via non-hermitian topology},}\ }\href {\doibase
  10.1038/s41377-024-01667-z} {\bibfield  {journal} {\bibinfo  {journal}
  {Light: Science \& Applications}\ }\textbf {\bibinfo {volume} {14}},\
  \bibinfo {pages} {6} (\bibinfo {year} {2025})}\BibitemShut {NoStop}%
\bibitem [{\citenamefont {Behrouzi}\ \emph {et~al.}(2025)\citenamefont
  {Behrouzi}, \citenamefont {Wu}, \citenamefont {Lin},\ and\ \citenamefont
  {Kante}}]{Behrouzi-25}%
  \BibitemOpen
  \bibfield  {author} {\bibinfo {author} {\bibfnamefont {Kamyar}\ \bibnamefont
  {Behrouzi}}, \bibinfo {author} {\bibfnamefont {Zhanni}\ \bibnamefont {Wu}},
  \bibinfo {author} {\bibfnamefont {Liwei}\ \bibnamefont {Lin}}, \ and\
  \bibinfo {author} {\bibfnamefont {Boubacar}\ \bibnamefont {Kante}},\
  }\bibfield  {title} {\enquote {\bibinfo {title} {Single plasmonic exceptional
  point nanoantenna coupled to a photonic integrated circuit sensor},}\ }\href
  {\doibase 10.1364/PRJ.540227} {\bibfield  {journal} {\bibinfo  {journal}
  {Photon. Res.}\ }\textbf {\bibinfo {volume} {13}},\ \bibinfo {pages}
  {632--641} (\bibinfo {year} {2025})}\BibitemShut {NoStop}%
\bibitem [{\citenamefont {Nag~Chowdhury}\ \emph {et~al.}(2025)\citenamefont
  {Nag~Chowdhury}, \citenamefont {Lahiri}, \citenamefont {Johnson},
  \citenamefont {De~La~Rue},\ and\ \citenamefont {Lahiri}}]{Chowdhury-25}%
  \BibitemOpen
  \bibfield  {author} {\bibinfo {author} {\bibfnamefont {Basudev}\ \bibnamefont
  {Nag~Chowdhury}}, \bibinfo {author} {\bibfnamefont {Pooja}\ \bibnamefont
  {Lahiri}}, \bibinfo {author} {\bibfnamefont {Nigel~P.}\ \bibnamefont
  {Johnson}}, \bibinfo {author} {\bibfnamefont {Richard~M.}\ \bibnamefont
  {De~La~Rue}}, \ and\ \bibinfo {author} {\bibfnamefont {Basudev}\ \bibnamefont
  {Lahiri}},\ }\bibfield  {title} {\enquote {\bibinfo {title}
  {Exceptional-point-enhanced superior sensing using asymmetric
  coupled-lossy-resonator based optical metasurface},}\ }\href {\doibase
  https://doi.org/10.1002/lpor.202401661} {\bibfield  {journal} {\bibinfo
  {journal} {Laser \& Photonics Reviews}\ }\textbf {\bibinfo {volume} {19}},\
  \bibinfo {pages} {2401661} (\bibinfo {year} {2025})}\BibitemShut {NoStop}%
\bibitem [{\citenamefont {Zheng}\ and\ \citenamefont {Chong}(2025)}]{Zheng-25}%
  \BibitemOpen
  \bibfield  {author} {\bibinfo {author} {\bibfnamefont {Xu}~\bibnamefont
  {Zheng}}\ and\ \bibinfo {author} {\bibfnamefont {Y.~D.}\ \bibnamefont
  {Chong}},\ }\bibfield  {title} {\enquote {\bibinfo {title} {Noise constraints
  for nonlinear exceptional point sensing},}\ }\href {\doibase
  10.1103/PhysRevLett.134.133801} {\bibfield  {journal} {\bibinfo  {journal}
  {Phys. Rev. Lett.}\ }\textbf {\bibinfo {volume} {134}},\ \bibinfo {pages}
  {133801} (\bibinfo {year} {2025})}\BibitemShut {NoStop}%
\bibitem [{\citenamefont {Wiersig}\ and\ \citenamefont
  {Rotter}(2026)}]{Wiersig-26}%
  \BibitemOpen
  \bibfield  {author} {\bibinfo {author} {\bibfnamefont {Jan}\ \bibnamefont
  {Wiersig}}\ and\ \bibinfo {author} {\bibfnamefont {Stefan}\ \bibnamefont
  {Rotter}},\ }\href {https://arxiv.org/abs/2603.10614} {\enquote {\bibinfo
  {title} {Fundamental limits of non-hermitian sensing from quantum fisher
  information},}\ } (\bibinfo {year} {2026}),\ \Eprint
  {http://arxiv.org/abs/2603.10614} {arXiv:2603.10614} \BibitemShut {NoStop}%
\bibitem [{\citenamefont {Tonielli}\ \emph {et~al.}(2020)\citenamefont
  {Tonielli}, \citenamefont {Budich}, \citenamefont {Altland},\ and\
  \citenamefont {Diehl}}]{Tonielli2020}%
  \BibitemOpen
  \bibfield  {author} {\bibinfo {author} {\bibfnamefont {F.}~\bibnamefont
  {Tonielli}}, \bibinfo {author} {\bibfnamefont {J.~C.}\ \bibnamefont
  {Budich}}, \bibinfo {author} {\bibfnamefont {A.}~\bibnamefont {Altland}}, \
  and\ \bibinfo {author} {\bibfnamefont {S.}~\bibnamefont {Diehl}},\ }\bibfield
   {title} {\enquote {\bibinfo {title} {Topological field theory far from
  equilibrium},}\ }\href {\doibase 10.1103/PhysRevLett.124.240404} {\bibfield
  {journal} {\bibinfo  {journal} {Phys. Rev. Lett.}\ }\textbf {\bibinfo
  {volume} {124}},\ \bibinfo {pages} {240404} (\bibinfo {year}
  {2020})}\BibitemShut {NoStop}%
\bibitem [{\citenamefont {Sayyad}\ \emph {et~al.}(2021)\citenamefont {Sayyad},
  \citenamefont {Yu}, \citenamefont {Grushin},\ and\ \citenamefont
  {Sieberer}}]{Sayyad2021}%
  \BibitemOpen
  \bibfield  {author} {\bibinfo {author} {\bibfnamefont {Sharareh}\
  \bibnamefont {Sayyad}}, \bibinfo {author} {\bibfnamefont {Jinlong}\
  \bibnamefont {Yu}}, \bibinfo {author} {\bibfnamefont {Adolfo~G.}\
  \bibnamefont {Grushin}}, \ and\ \bibinfo {author} {\bibfnamefont {Lukas~M.}\
  \bibnamefont {Sieberer}},\ }\bibfield  {title} {\enquote {\bibinfo {title}
  {Entanglement spectrum crossings reveal non-hermitian dynamical topology},}\
  }\href {\doibase 10.1103/PhysRevResearch.3.033022} {\bibfield  {journal}
  {\bibinfo  {journal} {Phys. Rev. Res.}\ }\textbf {\bibinfo {volume} {3}},\
  \bibinfo {pages} {033022} (\bibinfo {year} {2021})}\BibitemShut {NoStop}%
\bibitem [{\citenamefont {Starchl}\ and\ \citenamefont
  {Sieberer}(2022)}]{Starchl2022}%
  \BibitemOpen
  \bibfield  {author} {\bibinfo {author} {\bibfnamefont {Elias}\ \bibnamefont
  {Starchl}}\ and\ \bibinfo {author} {\bibfnamefont {Lukas~M.}\ \bibnamefont
  {Sieberer}},\ }\bibfield  {title} {\enquote {\bibinfo {title} {Relaxation to
  a parity-time symmetric generalized gibbs ensemble after a quantum quench in
  a driven-dissipative kitaev chain},}\ }\href {\doibase
  10.1103/PhysRevLett.129.220602} {\bibfield  {journal} {\bibinfo  {journal}
  {Phys. Rev. Lett.}\ }\textbf {\bibinfo {volume} {129}},\ \bibinfo {pages}
  {220602} (\bibinfo {year} {2022})}\BibitemShut {NoStop}%
\bibitem [{\citenamefont {Sayyad}\ \emph {et~al.}(2023)\citenamefont {Sayyad},
  \citenamefont {Stalhammar}, \citenamefont {Rødland},\ and\ \citenamefont
  {Kunst}}]{Sayyad2023}%
  \BibitemOpen
  \bibfield  {author} {\bibinfo {author} {\bibfnamefont {Sharareh}\
  \bibnamefont {Sayyad}}, \bibinfo {author} {\bibfnamefont {Marcus}\
  \bibnamefont {Stalhammar}}, \bibinfo {author} {\bibfnamefont {Lukas}\
  \bibnamefont {Rødland}}, \ and\ \bibinfo {author} {\bibfnamefont {Flore~K.}\
  \bibnamefont {Kunst}},\ }\bibfield  {title} {\enquote {\bibinfo {title}
  {{Symmetry-protected exceptional and nodal points in non-Hermitian
  systems}},}\ }\href {\doibase 10.21468/SciPostPhys.15.5.200} {\bibfield
  {journal} {\bibinfo  {journal} {SciPost Phys.}\ }\textbf {\bibinfo {volume}
  {15}},\ \bibinfo {pages} {200} (\bibinfo {year} {2023})}\BibitemShut
  {NoStop}%
\bibitem [{\citenamefont {Starchl}\ and\ \citenamefont
  {Sieberer}(2024)}]{Starchl2024}%
  \BibitemOpen
  \bibfield  {author} {\bibinfo {author} {\bibfnamefont {Elias}\ \bibnamefont
  {Starchl}}\ and\ \bibinfo {author} {\bibfnamefont {Lukas~M.}\ \bibnamefont
  {Sieberer}},\ }\bibfield  {title} {\enquote {\bibinfo {title} {Quantum
  quenches in driven-dissipative quadratic fermionic systems with parity-time
  symmetry},}\ }\href {\doibase 10.1103/PhysRevResearch.6.013016} {\bibfield
  {journal} {\bibinfo  {journal} {Phys. Rev. Res.}\ }\textbf {\bibinfo {volume}
  {6}},\ \bibinfo {pages} {013016} (\bibinfo {year} {2024})}\BibitemShut
  {NoStop}%
\bibitem [{\citenamefont {March\'e}\ \emph {et~al.}(2024)\citenamefont
  {March\'e}, \citenamefont {Yoshida}, \citenamefont {Nardin}, \citenamefont
  {Katsura},\ and\ \citenamefont {Mazza}}]{Marche2024}%
  \BibitemOpen
  \bibfield  {author} {\bibinfo {author} {\bibfnamefont {Alice}\ \bibnamefont
  {March\'e}}, \bibinfo {author} {\bibfnamefont {Hironobu}\ \bibnamefont
  {Yoshida}}, \bibinfo {author} {\bibfnamefont {Alberto}\ \bibnamefont
  {Nardin}}, \bibinfo {author} {\bibfnamefont {Hosho}\ \bibnamefont {Katsura}},
  \ and\ \bibinfo {author} {\bibfnamefont {Leonardo}\ \bibnamefont {Mazza}},\
  }\bibfield  {title} {\enquote {\bibinfo {title} {Universality and two-body
  losses: Lessons from the effective non-hermitian dynamics of two
  particles},}\ }\href {\doibase 10.1103/PhysRevA.110.033321} {\bibfield
  {journal} {\bibinfo  {journal} {Phys. Rev. A}\ }\textbf {\bibinfo {volume}
  {110}},\ \bibinfo {pages} {033321} (\bibinfo {year} {2024})}\BibitemShut
  {NoStop}%
\bibitem [{\citenamefont {Jiang}\ \emph {et~al.}(2019)\citenamefont {Jiang},
  \citenamefont {Mei}, \citenamefont {Zuo}, \citenamefont {Zhai}, \citenamefont
  {Li}, \citenamefont {Wen},\ and\ \citenamefont {Du}}]{Jian2019}%
  \BibitemOpen
  \bibfield  {author} {\bibinfo {author} {\bibfnamefont {Yue}\ \bibnamefont
  {Jiang}}, \bibinfo {author} {\bibfnamefont {Yefeng}\ \bibnamefont {Mei}},
  \bibinfo {author} {\bibfnamefont {Ying}\ \bibnamefont {Zuo}}, \bibinfo
  {author} {\bibfnamefont {Yanhua}\ \bibnamefont {Zhai}}, \bibinfo {author}
  {\bibfnamefont {Jensen}\ \bibnamefont {Li}}, \bibinfo {author} {\bibfnamefont
  {Jianming}\ \bibnamefont {Wen}}, \ and\ \bibinfo {author} {\bibfnamefont
  {Shengwang}\ \bibnamefont {Du}},\ }\bibfield  {title} {\enquote {\bibinfo
  {title} {Anti-parity-time symmetric optical four-wave mixing in cold
  atoms},}\ }\href {\doibase 10.1103/PhysRevLett.123.193604} {\bibfield
  {journal} {\bibinfo  {journal} {Phys. Rev. Lett.}\ }\textbf {\bibinfo
  {volume} {123}},\ \bibinfo {pages} {193604} (\bibinfo {year}
  {2019})}\BibitemShut {NoStop}%
\bibitem [{\citenamefont {Kawabata}\ \emph {et~al.}(2021)\citenamefont
  {Kawabata}, \citenamefont {Shiozaki},\ and\ \citenamefont
  {Ryu}}]{Kawabata2021}%
  \BibitemOpen
  \bibfield  {author} {\bibinfo {author} {\bibfnamefont {Kohei}\ \bibnamefont
  {Kawabata}}, \bibinfo {author} {\bibfnamefont {Ken}\ \bibnamefont
  {Shiozaki}}, \ and\ \bibinfo {author} {\bibfnamefont {Shinsei}\ \bibnamefont
  {Ryu}},\ }\bibfield  {title} {\enquote {\bibinfo {title} {Topological field
  theory of non-hermitian systems},}\ }\href {\doibase
  10.1103/PhysRevLett.126.216405} {\bibfield  {journal} {\bibinfo  {journal}
  {Phys. Rev. Lett.}\ }\textbf {\bibinfo {volume} {126}},\ \bibinfo {pages}
  {216405} (\bibinfo {year} {2021})}\BibitemShut {NoStop}%
\bibitem [{\citenamefont {Sayyad}\ \emph {et~al.}(2022)\citenamefont {Sayyad},
  \citenamefont {Hannukainen},\ and\ \citenamefont {Grushin}}]{Sayyad2022c}%
  \BibitemOpen
  \bibfield  {author} {\bibinfo {author} {\bibfnamefont {Sharareh}\
  \bibnamefont {Sayyad}}, \bibinfo {author} {\bibfnamefont {Julia~D.}\
  \bibnamefont {Hannukainen}}, \ and\ \bibinfo {author} {\bibfnamefont
  {Adolfo~G.}\ \bibnamefont {Grushin}},\ }\bibfield  {title} {\enquote
  {\bibinfo {title} {Non-hermitian chiral anomalies},}\ }\href {\doibase
  10.1103/PhysRevResearch.4.L042004} {\bibfield  {journal} {\bibinfo  {journal}
  {Phys. Rev. Res.}\ }\textbf {\bibinfo {volume} {4}},\ \bibinfo {pages}
  {L042004} (\bibinfo {year} {2022})}\BibitemShut {NoStop}%
\bibitem [{\citenamefont {Sayyad}\ and\ \citenamefont {Lado}(2023)}]{Lado2023}%
  \BibitemOpen
  \bibfield  {author} {\bibinfo {author} {\bibfnamefont {Sharareh}\
  \bibnamefont {Sayyad}}\ and\ \bibinfo {author} {\bibfnamefont {Jose~L.}\
  \bibnamefont {Lado}},\ }\bibfield  {title} {\enquote {\bibinfo {title}
  {Topological phase diagrams of exactly solvable non-hermitian interacting
  kitaev chains},}\ }\href {\doibase 10.1103/PhysRevResearch.5.L022046}
  {\bibfield  {journal} {\bibinfo  {journal} {Phys. Rev. Res.}\ }\textbf
  {\bibinfo {volume} {5}},\ \bibinfo {pages} {L022046} (\bibinfo {year}
  {2023})}\BibitemShut {NoStop}%
\bibitem [{\citenamefont {Yoshida}\ \emph {et~al.}(2024)\citenamefont
  {Yoshida}, \citenamefont {Zhang}, \citenamefont {Neupert},\ and\
  \citenamefont {Kawakami}}]{Yoshida2024}%
  \BibitemOpen
  \bibfield  {author} {\bibinfo {author} {\bibfnamefont {Tsuneya}\ \bibnamefont
  {Yoshida}}, \bibinfo {author} {\bibfnamefont {Song-Bo}\ \bibnamefont
  {Zhang}}, \bibinfo {author} {\bibfnamefont {Titus}\ \bibnamefont {Neupert}},
  \ and\ \bibinfo {author} {\bibfnamefont {Norio}\ \bibnamefont {Kawakami}},\
  }\bibfield  {title} {\enquote {\bibinfo {title} {Non-hermitian mott skin
  effect},}\ }\href {\doibase 10.1103/PhysRevLett.133.076502} {\bibfield
  {journal} {\bibinfo  {journal} {Phys. Rev. Lett.}\ }\textbf {\bibinfo
  {volume} {133}},\ \bibinfo {pages} {076502} (\bibinfo {year}
  {2024})}\BibitemShut {NoStop}%
\bibitem [{\citenamefont {Sayyad}(2024)}]{Sayyad2024d}%
  \BibitemOpen
  \bibfield  {author} {\bibinfo {author} {\bibfnamefont {Sharareh}\
  \bibnamefont {Sayyad}},\ }\bibfield  {title} {\enquote {\bibinfo {title}
  {Non-hermitian chiral anomalies in interacting systems},}\ }\href {\doibase
  10.1103/PhysRevResearch.6.L012028} {\bibfield  {journal} {\bibinfo  {journal}
  {Phys. Rev. Res.}\ }\textbf {\bibinfo {volume} {6}},\ \bibinfo {pages}
  {L012028} (\bibinfo {year} {2024})}\BibitemShut {NoStop}%
\bibitem [{\citenamefont {Le~Gal}\ \emph {et~al.}(2024)\citenamefont {Le~Gal},
  \citenamefont {Turkeshi},\ and\ \citenamefont {Schir\`o}}]{Legal2024}%
  \BibitemOpen
  \bibfield  {author} {\bibinfo {author} {\bibfnamefont {Youenn}\ \bibnamefont
  {Le~Gal}}, \bibinfo {author} {\bibfnamefont {Xhek}\ \bibnamefont {Turkeshi}},
  \ and\ \bibinfo {author} {\bibfnamefont {Marco}\ \bibnamefont {Schir\`o}},\
  }\bibfield  {title} {\enquote {\bibinfo {title} {Entanglement dynamics in
  monitored systems and the role of quantum jumps},}\ }\href {\doibase
  10.1103/PRXQuantum.5.030329} {\bibfield  {journal} {\bibinfo  {journal} {PRX
  Quantum}\ }\textbf {\bibinfo {volume} {5}},\ \bibinfo {pages} {030329}
  (\bibinfo {year} {2024})}\BibitemShut {NoStop}%
\bibitem [{Note1()}]{Note1}%
  \BibitemOpen
  \bibinfo {note} {We note that in this case the geometric multiplicity of the
  eigenvalue is one.}\BibitemShut {Stop}%
\bibitem [{\citenamefont {Bid}\ and\ \citenamefont
  {Schomerus}(2025{\natexlab{a}})}]{bid_PhysRevResearch}%
  \BibitemOpen
  \bibfield  {author} {\bibinfo {author} {\bibfnamefont {Subhajyoti}\
  \bibnamefont {Bid}}\ and\ \bibinfo {author} {\bibfnamefont {Henning}\
  \bibnamefont {Schomerus}},\ }\bibfield  {title} {\enquote {\bibinfo {title}
  {Uniform response theory of non-hermitian systems: Non-hermitian physics
  beyond the exceptional point},}\ }\href {\doibase
  10.1103/PhysRevResearch.7.023062} {\bibfield  {journal} {\bibinfo  {journal}
  {Phys. Rev. Res.}\ }\textbf {\bibinfo {volume} {7}},\ \bibinfo {pages}
  {023062} (\bibinfo {year} {2025}{\natexlab{a}})}\BibitemShut {NoStop}%
\bibitem [{\citenamefont {Bid}\ and\ \citenamefont
  {Schomerus}(2025{\natexlab{b}})}]{bid_2025}%
  \BibitemOpen
  \bibfield  {author} {\bibinfo {author} {\bibfnamefont {Subhajyoti}\
  \bibnamefont {Bid}}\ and\ \bibinfo {author} {\bibfnamefont {Henning}\
  \bibnamefont {Schomerus}},\ }\href {https://arxiv.org/abs/2507.22158}
  {\enquote {\bibinfo {title} {Fragmented exceptional points and their bulk and
  edge realizations in lattice models},}\ } (\bibinfo {year}
  {2025}{\natexlab{b}}),\ \Eprint {http://arxiv.org/abs/2507.22158}
  {arXiv:2507.22158} \BibitemShut {NoStop}%
\bibitem [{\citenamefont {Shiralieva}\ \emph {et~al.}(2025)\citenamefont
  {Shiralieva}, \citenamefont {Starkov},\ and\ \citenamefont
  {Trauzettel}}]{shiralieva2025}%
  \BibitemOpen
  \bibfield  {author} {\bibinfo {author} {\bibfnamefont {Aysel}\ \bibnamefont
  {Shiralieva}}, \bibinfo {author} {\bibfnamefont {Grigory~A.}\ \bibnamefont
  {Starkov}}, \ and\ \bibinfo {author} {\bibfnamefont {Björn}\ \bibnamefont
  {Trauzettel}},\ }\href {https://arxiv.org/abs/2509.11856} {\enquote {\bibinfo
  {title} {Multi-block exceptional points in open quantum systems},}\ }
  (\bibinfo {year} {2025}),\ \Eprint {http://arxiv.org/abs/2509.11856}
  {arXiv:2509.11856} \BibitemShut {NoStop}%
\bibitem [{\citenamefont {Wiersig}\ and\ \citenamefont
  {Chen}(2025)}]{Wiersig2025}%
  \BibitemOpen
  \bibfield  {author} {\bibinfo {author} {\bibfnamefont {Jan}\ \bibnamefont
  {Wiersig}}\ and\ \bibinfo {author} {\bibfnamefont {Weijian}\ \bibnamefont
  {Chen}},\ }\bibfield  {title} {\enquote {\bibinfo {title} {Higher-order
  exceptional points in composite non-hermitian systems},}\ }\href {\doibase
  10.1103/tmw5-fcph} {\bibfield  {journal} {\bibinfo  {journal} {Phys. Rev.
  Res.}\ }\textbf {\bibinfo {volume} {7}},\ \bibinfo {pages} {033034} (\bibinfo
  {year} {2025})}\BibitemShut {NoStop}%
\bibitem [{\citenamefont {Sayyad}\ and\ \citenamefont
  {Starkov}(2026)}]{Sayyad2026}%
  \BibitemOpen
  \bibfield  {author} {\bibinfo {author} {\bibfnamefont {Sharareh}\
  \bibnamefont {Sayyad}}\ and\ \bibinfo {author} {\bibfnamefont {Grigory~A.}\
  \bibnamefont {Starkov}},\ }\bibfield  {title} {\enquote {\bibinfo {title}
  {Characterizing all non-hermitian degeneracies using algebraic approaches:
  Defectiveness and asymptotic behavior},}\ }\href
  {https://arxiv.org/abs/2604.16140} {\bibfield  {journal} {\bibinfo  {journal}
  {arXiv:2604.16140}\ } (\bibinfo {year} {2026})}\BibitemShut {NoStop}%
\bibitem [{\citenamefont {Lidskii}(1966)}]{Lidskii-1966}%
  \BibitemOpen
  \bibfield  {author} {\bibinfo {author} {\bibfnamefont {V.B.}\ \bibnamefont
  {Lidskii}},\ }\bibfield  {title} {\enquote {\bibinfo {title} {Perturbation
  theory of non-conjugate operators},}\ }\href {\doibase
  10.1016/0041-5553(66)90033-4} {\bibfield  {journal} {\bibinfo  {journal}
  {USSR Computational Mathematics and Mathematical Physics}\ }\textbf {\bibinfo
  {volume} {6}},\ \bibinfo {pages} {73–85} (\bibinfo {year}
  {1966})}\BibitemShut {NoStop}%
\bibitem [{\citenamefont {Moro}\ and\ \citenamefont
  {Dopico}(2003)}]{Moro-2003}%
  \BibitemOpen
  \bibfield  {author} {\bibinfo {author} {\bibfnamefont {Julio}\ \bibnamefont
  {Moro}}\ and\ \bibinfo {author} {\bibfnamefont {Froil{\'a}n~M.}\ \bibnamefont
  {Dopico}},\ }\enquote {\bibinfo {title} {First order eigenvalue perturbation
  theory and the newton diagram},}\ in\ \href {\doibase
  10.1007/978-1-4757-4532-0_6} {\emph {\bibinfo {booktitle} {Applied
  Mathematics and Scientific Computing}}},\ \bibinfo {editor} {edited by\
  \bibinfo {editor} {\bibfnamefont {Zlatko}\ \bibnamefont {Drma{\v{c}}}},
  \bibinfo {editor} {\bibfnamefont {Vjeran}\ \bibnamefont {Hari}}, \bibinfo
  {editor} {\bibfnamefont {Luka}\ \bibnamefont {Sopta}}, \bibinfo {editor}
  {\bibfnamefont {Zvonimir}\ \bibnamefont {Tutek}}, \ and\ \bibinfo {editor}
  {\bibfnamefont {Kre{\v{s}}imir}\ \bibnamefont {Veseli{\'{c}}}}}\ (\bibinfo
  {publisher} {Springer US},\ \bibinfo {address} {Boston, MA},\ \bibinfo {year}
  {2003})\ pp.\ \bibinfo {pages} {143--175}\BibitemShut {NoStop}%
\bibitem [{Note2()}]{Note2}%
  \BibitemOpen
  \bibinfo {note} {We note that varying the trace merely adds a constant shift
  to the eigenvalues and does not affect the structure of
  degeneracies.}\BibitemShut {Stop}%
\bibitem [{\citenamefont {Arnold}(1971)}]{Arnold1971}%
  \BibitemOpen
  \bibfield  {author} {\bibinfo {author} {\bibfnamefont {V~I}\ \bibnamefont
  {Arnold}},\ }\bibfield  {title} {\enquote {\bibinfo {title} {On matrices
  depending on parameters},}\ }\href {\doibase 10.1070/rm1971v026n02abeh003827}
  {\bibfield  {journal} {\bibinfo  {journal} {Russian Mathematical Surveys}\
  }\textbf {\bibinfo {volume} {26}},\ \bibinfo {pages} {29--43} (\bibinfo
  {year} {1971})}\BibitemShut {NoStop}%
\bibitem [{\citenamefont {Gantmakher}(1998)}]{Gantmakher1998}%
  \BibitemOpen
  \bibfield  {author} {\bibinfo {author} {\bibfnamefont {F.~R.}\ \bibnamefont
  {Gantmakher}},\ }\href@noop {} {\emph {\bibinfo {title} {The theory of
  matrices, volume 1}}},\ AMS Chelsea Publishing\ (\bibinfo  {publisher}
  {American Mathematical Society},\ \bibinfo {address} {Providence, RI},\
  \bibinfo {year} {1998})\BibitemShut {NoStop}%
\bibitem [{\citenamefont {Sayyad}\ and\ \citenamefont
  {Kunst}(2022)}]{Sayyad2022}%
  \BibitemOpen
  \bibfield  {author} {\bibinfo {author} {\bibfnamefont {Sharareh}\
  \bibnamefont {Sayyad}}\ and\ \bibinfo {author} {\bibfnamefont {Flore~K.}\
  \bibnamefont {Kunst}},\ }\bibfield  {title} {\enquote {\bibinfo {title}
  {Realizing exceptional points of any order in the presence of symmetry},}\
  }\href {\doibase 10.1103/PhysRevResearch.4.023130} {\bibfield  {journal}
  {\bibinfo  {journal} {Phys. Rev. Res.}\ }\textbf {\bibinfo {volume} {4}},\
  \bibinfo {pages} {023130} (\bibinfo {year} {2022})}\BibitemShut {NoStop}%
\bibitem [{\citenamefont {Mandal}\ and\ \citenamefont
  {Bergholtz}(2021)}]{Bergholtz-21}%
  \BibitemOpen
  \bibfield  {author} {\bibinfo {author} {\bibfnamefont {Ipsita}\ \bibnamefont
  {Mandal}}\ and\ \bibinfo {author} {\bibfnamefont {Emil~J.}\ \bibnamefont
  {Bergholtz}},\ }\bibfield  {title} {\enquote {\bibinfo {title} {Symmetry and
  higher-order exceptional points},}\ }\href {\doibase
  10.1103/PhysRevLett.127.186601} {\bibfield  {journal} {\bibinfo  {journal}
  {Phys. Rev. Lett.}\ }\textbf {\bibinfo {volume} {127}},\ \bibinfo {pages}
  {186601} (\bibinfo {year} {2021})}\BibitemShut {NoStop}%
\bibitem [{Note3()}]{Note3}%
  \BibitemOpen
  \bibinfo {note} {This is a direct consequence of the rank-nullity
  theorem.}\BibitemShut {Stop}%
\bibitem [{Note4()}]{Note4}%
  \BibitemOpen
  \bibinfo {note} {One should keep in mind that the lines themselves do not
  belong to the EP3 manifold. To get the intersection, we need to extend the
  sheets by their closure.}\BibitemShut {Stop}%
\bibitem [{Note5()}]{Note5}%
  \BibitemOpen
  \bibinfo {note} {A similarity transformation $A\rightarrow SAS^{-1}$ with
  fixed $S$ is a bounded linear operator on the space of matrices, hence it
  commutes with taking the limit.}\BibitemShut {Stop}%
\bibitem [{Note6()}]{Note6}%
  \BibitemOpen
  \bibinfo {note} {Note that $\protect \overline {A}$ stands for the closure of
  set $A$.}\BibitemShut {Stop}%
\bibitem [{\citenamefont {Gerstenhaber}(1959)}]{gerstenhaber_1959}%
  \BibitemOpen
  \bibfield  {author} {\bibinfo {author} {\bibfnamefont {Murray}\ \bibnamefont
  {Gerstenhaber}},\ }\bibfield  {title} {\enquote {\bibinfo {title} {On
  {Nilalgebras} and {Linear} {Varieties} of {Nilpotent} {Matrices}, {III}},}\
  }\href {\doibase 10.2307/1969896} {\bibfield  {journal} {\bibinfo  {journal}
  {Annals of Mathematics}\ }\textbf {\bibinfo {volume} {70}},\ \bibinfo {pages}
  {167--205} (\bibinfo {year} {1959})},\ \bibinfo {note} {publisher: [Annals of
  Mathematics, Trustees of Princeton University on Behalf of the Annals of
  Mathematics, Mathematics Department, Princeton University]}\BibitemShut
  {NoStop}%
\bibitem [{\citenamefont {Kraft}\ and\ \citenamefont
  {Procesi}(1982)}]{kraft_1982}%
  \BibitemOpen
  \bibfield  {author} {\bibinfo {author} {\bibfnamefont {Hanspeter}\
  \bibnamefont {Kraft}}\ and\ \bibinfo {author} {\bibfnamefont {Claudio}\
  \bibnamefont {Procesi}},\ }\bibfield  {title} {\enquote {\bibinfo {title} {On
  the geometry of conjugacy classes in classical groups},}\ }\href {\doibase
  10.1007/BF02565876} {\bibfield  {journal} {\bibinfo  {journal} {Commentarii
  Mathematici Helvetici}\ }\textbf {\bibinfo {volume} {57}},\ \bibinfo {pages}
  {539--602} (\bibinfo {year} {1982})}\BibitemShut {NoStop}%
\bibitem [{Note7()}]{Note7}%
  \BibitemOpen
  \bibinfo {note} {These structures are also known as the Young diagram
  lattices on the partition of an integer~\cite {Latapy2009}.}\BibitemShut
  {Stop}%
\bibitem [{\citenamefont {Latapy}\ and\ \citenamefont
  {Phan}(2009)}]{Latapy2009}%
  \BibitemOpen
  \bibfield  {author} {\bibinfo {author} {\bibfnamefont {Matthieu}\
  \bibnamefont {Latapy}}\ and\ \bibinfo {author} {\bibfnamefont {Thi Ha~Duong}\
  \bibnamefont {Phan}},\ }\bibfield  {title} {\enquote {\bibinfo {title} {The
  lattice of integer partitions and its infinite extension},}\ }\href {\doibase
  10.1016/j.disc.2008.02.002} {\bibfield  {journal} {\bibinfo  {journal}
  {Discrete Mathematics}\ }\textbf {\bibinfo {volume} {309}},\ \bibinfo {pages}
  {1357–1367} (\bibinfo {year} {2009})}\BibitemShut {NoStop}%
\bibitem [{\citenamefont
  {Mostafazadeh}(2002{\natexlab{a}})}]{mostafazadeh2002pseudo}%
  \BibitemOpen
  \bibfield  {author} {\bibinfo {author} {\bibfnamefont {Ali}\ \bibnamefont
  {Mostafazadeh}},\ }\bibfield  {title} {\enquote {\bibinfo {title}
  {{Pseudo-Hermiticity versus PT symmetry: The necessary condition for the
  reality of the spectrum of a non-Hermitian Hamiltonian}},}\ }\href {\doibase
  10.1063/1.1418246} {\bibfield  {journal} {\bibinfo  {journal} {Journal of
  Mathematical Physics}\ }\textbf {\bibinfo {volume} {43}},\ \bibinfo {pages}
  {205--214} (\bibinfo {year} {2002}{\natexlab{a}})}\BibitemShut {NoStop}%
\bibitem [{\citenamefont {Mostafazadeh}(2002{\natexlab{b}})}]{mostafazadeh2}%
  \BibitemOpen
  \bibfield  {author} {\bibinfo {author} {\bibfnamefont {Ali}\ \bibnamefont
  {Mostafazadeh}},\ }\bibfield  {title} {\enquote {\bibinfo {title}
  {{Pseudo-Hermiticity versus PT-symmetry. II. A complete characterization of
  non-Hermitian Hamiltonians with a real spectrum}},}\ }\href {\doibase
  10.1063/1.1461427} {\bibfield  {journal} {\bibinfo  {journal} {Journal of
  Mathematical Physics}\ }\textbf {\bibinfo {volume} {43}},\ \bibinfo {pages}
  {2814--2816} (\bibinfo {year} {2002}{\natexlab{b}})}\BibitemShut {NoStop}%
\bibitem [{\citenamefont {Mostafazadeh}(2002{\natexlab{c}})}]{mostafazadeh3}%
  \BibitemOpen
  \bibfield  {author} {\bibinfo {author} {\bibfnamefont {Ali}\ \bibnamefont
  {Mostafazadeh}},\ }\bibfield  {title} {\enquote {\bibinfo {title}
  {{Pseudo-Hermiticity versus PT-symmetry III: Equivalence of
  pseudo-Hermiticity and the presence of antilinear symmetries}},}\ }\href
  {\doibase 10.1063/1.1489072} {\bibfield  {journal} {\bibinfo  {journal}
  {Journal of Mathematical Physics}\ }\textbf {\bibinfo {volume} {43}},\
  \bibinfo {pages} {3944--3951} (\bibinfo {year}
  {2002}{\natexlab{c}})}\BibitemShut {NoStop}%
\bibitem [{\citenamefont {Mostafazadeh}(2010)}]{mostafazadeh2010}%
  \BibitemOpen
  \bibfield  {author} {\bibinfo {author} {\bibfnamefont {Ali}\ \bibnamefont
  {Mostafazadeh}},\ }\bibfield  {title} {\enquote {\bibinfo {title}
  {{Pseudo}-{Hermitian} {Representation} {of} {Quantum} {Mechanics}},}\ }\href
  {\doibase 10.1142/S0219887810004816} {\bibfield  {journal} {\bibinfo
  {journal} {International Journal of Geometric Methods in Modern Physics}\
  }\textbf {\bibinfo {volume} {07}},\ \bibinfo {pages} {1191--1306} (\bibinfo
  {year} {2010})}\BibitemShut {NoStop}%
\bibitem [{\citenamefont {Zhang}\ \emph {et~al.}(2020)\citenamefont {Zhang},
  \citenamefont {Qin},\ and\ \citenamefont {Xiao}}]{ZhangQinXiao-20}%
  \BibitemOpen
  \bibfield  {author} {\bibinfo {author} {\bibfnamefont {Ruili}\ \bibnamefont
  {Zhang}}, \bibinfo {author} {\bibfnamefont {Hong}\ \bibnamefont {Qin}}, \
  and\ \bibinfo {author} {\bibfnamefont {Jianyuan}\ \bibnamefont {Xiao}},\
  }\bibfield  {title} {\enquote {\bibinfo {title} {{PT}-symmetry entails
  pseudo-hermiticity regardless of diagonalizability},}\ }\href {\doibase
  10.1063/1.5117211} {\bibfield  {journal} {\bibinfo  {journal} {Journal of
  Mathematical Physics}\ }\textbf {\bibinfo {volume} {61}},\ \bibinfo {pages}
  {012101} (\bibinfo {year} {2020})}\BibitemShut {NoStop}%
\bibitem [{\citenamefont {S\'a}\ \emph {et~al.}(2023)\citenamefont {S\'a},
  \citenamefont {Ribeiro},\ and\ \citenamefont {Prosen}}]{Sa_2023}%
  \BibitemOpen
  \bibfield  {author} {\bibinfo {author} {\bibfnamefont {Lucas}\ \bibnamefont
  {S\'a}}, \bibinfo {author} {\bibfnamefont {Pedro}\ \bibnamefont {Ribeiro}}, \
  and\ \bibinfo {author} {\bibfnamefont {Tomas}\ \bibnamefont {Prosen}},\
  }\bibfield  {title} {\enquote {\bibinfo {title} {Symmetry classification of
  many-body {Lindbladians}: Tenfold way and beyond},}\ }\href {\doibase
  10.1103/PhysRevX.13.031019} {\bibfield  {journal} {\bibinfo  {journal} {Phys.
  Rev. X}\ }\textbf {\bibinfo {volume} {13}},\ \bibinfo {pages} {031019}
  (\bibinfo {year} {2023})}\BibitemShut {NoStop}%
\bibitem [{Note8()}]{Note8}%
  \BibitemOpen
  \bibinfo {note} {A Hermitian operator $\eta $ can be diagonalized by a
  unitary transformation: $\eta = U^\dagger D U$, where $U$ is an invertible
  matrix and $D = \protect \diag {(\varepsilon _1, \varepsilon _2,...)}$ with
  real eigenvalues $\varepsilon _i$. Without the loss of generality let's
  assume that the eigenvalues are ordered $\varepsilon _i\geqslant \varepsilon
  _i+1$, first $p$ eigenvalues are positive and the rest $q$ eigenvalues are
  negative. Then $\eta = S_0^\dagger \eta _{p,q}S_0$ with $S_0 = |D|^{1/2}U$,
  where $|D|^{1/2} = \protect \diag {(\protect \sqrt {|\varepsilon _1|},
  \protect \sqrt {|\varepsilon _2|},...)}$}\BibitemShut {NoStop}%
\bibitem [{\citenamefont {Mostafazadeh}(2004)}]{Mostafazadeh2004}%
  \BibitemOpen
  \bibfield  {author} {\bibinfo {author} {\bibfnamefont {Ali}\ \bibnamefont
  {Mostafazadeh}},\ }\bibfield  {title} {\enquote {\bibinfo {title}
  {Pseudounitary operators and pseudounitary quantum dynamics},}\ }\href
  {\doibase 10.1063/1.1646448} {\bibfield  {journal} {\bibinfo  {journal}
  {Journal of Mathematical Physics}\ }\textbf {\bibinfo {volume} {45}},\
  \bibinfo {pages} {932–946} (\bibinfo {year} {2004})}\BibitemShut {NoStop}%
\bibitem [{\citenamefont {Gohberg}\ \emph {et~al.}(2006)\citenamefont
  {Gohberg}, \citenamefont {Lancaster},\ and\ \citenamefont
  {Rodman}}]{Gohberg2006}%
  \BibitemOpen
  \bibfield  {author} {\bibinfo {author} {\bibfnamefont {Israel}\ \bibnamefont
  {Gohberg}}, \bibinfo {author} {\bibfnamefont {Peter}\ \bibnamefont
  {Lancaster}}, \ and\ \bibinfo {author} {\bibfnamefont {Leiba}\ \bibnamefont
  {Rodman}},\ }\href@noop {} {\emph {\bibinfo {title} {Indefinite linear
  algebra and applications}}}\ (\bibinfo  {publisher} {Springer Science \&
  Business Media},\ \bibinfo {year} {2006})\BibitemShut {NoStop}%
\bibitem [{Note9()}]{Note9}%
  \BibitemOpen
  \bibinfo {note} {See Theorem 5.1.1 in Ref.\cite {Gohberg2006}.}\BibitemShut
  {Stop}%
\bibitem [{Note10()}]{Note10}%
  \BibitemOpen
  \bibinfo {note} {This matrix is also known as the standard involutory
  permutation~(sip) matrix, or the reversal matrix~\cite
  {horn2012matrix}.}\BibitemShut {Stop}%
\bibitem [{Note11()}]{Note11}%
  \BibitemOpen
  \bibinfo {note} {For example, it follows from Theorem 9.1.1 in Ref.~\cite
  {Gohberg2006}}\BibitemShut {NoStop}%
\bibitem [{\citenamefont {Krein}(1950)}]{Krein-1950}%
  \BibitemOpen
  \bibfield  {author} {\bibinfo {author} {\bibfnamefont {MG}~\bibnamefont
  {Krein}},\ }\bibfield  {title} {\enquote {\bibinfo {title} {A generalization
  of some investigations of linear differential equations with periodic
  coefficients},}\ }in\ \href@noop {} {\emph {\bibinfo {booktitle} {Doklady
  Akad. Nauk SSSR}}},\ Vol.~\bibinfo {volume} {73}\ (\bibinfo {year} {1950})\
  pp.\ \bibinfo {pages} {445--448}\BibitemShut {NoStop}%
\bibitem [{\citenamefont {Gel'fand}\ and\ \citenamefont
  {Lidskii}(1955)}]{GelfandLidskii-1955}%
  \BibitemOpen
  \bibfield  {author} {\bibinfo {author} {\bibfnamefont {Izrail~Moiseevich}\
  \bibnamefont {Gel'fand}}\ and\ \bibinfo {author} {\bibfnamefont
  {Viktor~Borisovich}\ \bibnamefont {Lidskii}},\ }\bibfield  {title} {\enquote
  {\bibinfo {title} {On the structure of the regions of stability of linear
  canonical systems of differential equations with periodic coefficients},}\
  }\href
  {https://www.mathnet.ru/php/archive.phtml?wshow=paper&jrnid=rm&paperid=7943&option_lang=eng}
  {\bibfield  {journal} {\bibinfo  {journal} {Uspekhi Matematicheskikh Nauk}\
  }\textbf {\bibinfo {volume} {10}},\ \bibinfo {pages} {3--40} (\bibinfo {year}
  {1955})}\BibitemShut {NoStop}%
\bibitem [{\citenamefont {Melkani}(2023)}]{Melkani-23}%
  \BibitemOpen
  \bibfield  {author} {\bibinfo {author} {\bibfnamefont {Abhijeet}\
  \bibnamefont {Melkani}},\ }\bibfield  {title} {\enquote {\bibinfo {title}
  {Degeneracies and symmetry breaking in pseudo-hermitian matrices},}\ }\href
  {\doibase 10.1103/PhysRevResearch.5.023035} {\bibfield  {journal} {\bibinfo
  {journal} {Phys. Rev. Res.}\ }\textbf {\bibinfo {volume} {5}},\ \bibinfo
  {pages} {023035} (\bibinfo {year} {2023})}\BibitemShut {NoStop}%
\bibitem [{\citenamefont {Starkov}\ \emph
  {et~al.}(2023{\natexlab{a}})\citenamefont {Starkov}, \citenamefont {Fistul},\
  and\ \citenamefont {Eremin}}]{StarkovFistulEremin-23}%
  \BibitemOpen
  \bibfield  {author} {\bibinfo {author} {\bibfnamefont {Grigory~A.}\
  \bibnamefont {Starkov}}, \bibinfo {author} {\bibfnamefont {Mikhail~V.}\
  \bibnamefont {Fistul}}, \ and\ \bibinfo {author} {\bibfnamefont {Ilya~M.}\
  \bibnamefont {Eremin}},\ }\bibfield  {title} {\enquote {\bibinfo {title}
  {Formation of exceptional points in pseudo-hermitian systems},}\ }\href
  {\doibase 10.1103/PhysRevA.108.022206} {\bibfield  {journal} {\bibinfo
  {journal} {Phys. Rev. A}\ }\textbf {\bibinfo {volume} {108}},\ \bibinfo
  {pages} {022206} (\bibinfo {year} {2023}{\natexlab{a}})}\BibitemShut
  {NoStop}%
\bibitem [{\citenamefont {Starkov}\ \emph
  {et~al.}(2023{\natexlab{b}})\citenamefont {Starkov}, \citenamefont {Fistul},\
  and\ \citenamefont {Eremin}}]{StarkovFistulEremin-23b}%
  \BibitemOpen
  \bibfield  {author} {\bibinfo {author} {\bibfnamefont {Grigory~A.}\
  \bibnamefont {Starkov}}, \bibinfo {author} {\bibfnamefont {Mikhail~V.}\
  \bibnamefont {Fistul}}, \ and\ \bibinfo {author} {\bibfnamefont {Ilya~M.}\
  \bibnamefont {Eremin}},\ }\bibfield  {title} {\enquote {\bibinfo {title}
  {Schrieffer-wolff transformation for non-hermitian systems: Application for
  $\mathcal{PT}$-symmetric circuit qed},}\ }\href {\doibase
  10.1103/PhysRevB.108.235417} {\bibfield  {journal} {\bibinfo  {journal}
  {Phys. Rev. B}\ }\textbf {\bibinfo {volume} {108}},\ \bibinfo {pages}
  {235417} (\bibinfo {year} {2023}{\natexlab{b}})}\BibitemShut {NoStop}%
\bibitem [{\citenamefont {Starkov}(2024)}]{Starkov-24}%
  \BibitemOpen
  \bibfield  {author} {\bibinfo {author} {\bibfnamefont {Grigory~A.}\
  \bibnamefont {Starkov}},\ }\href {https://arxiv.org/abs/2402.07690} {\enquote
  {\bibinfo {title} {Interplay of pseudo-hermitian symmetries and degenerate
  manifolds in the eigenspectrum of non-hermitian systems},}\ } (\bibinfo
  {year} {2024}),\ \Eprint {http://arxiv.org/abs/2402.07690} {arXiv:2402.07690}
  \BibitemShut {NoStop}%
\bibitem [{Note12()}]{Note12}%
  \BibitemOpen
  \bibinfo {note} {There is no contradiction here. Pseudo-Hermitian symmetry is
  still satisfied in the sense that the corresponding degeneracy at the complex
  conjugate energy behaves symmetrically.}\BibitemShut {Stop}%
\bibitem [{\citenamefont {Djokovi{\'c}}(2006)}]{Djokovic2006}%
  \BibitemOpen
  \bibfield  {author} {\bibinfo {author} {\bibfnamefont {Dragomir~{\v{Z}}}\
  \bibnamefont {Djokovi{\'c}}},\ }\bibfield  {title} {\enquote {\bibinfo
  {title} {Closures of conjugacy classes in classical real linear lie
  groups},}\ }in\ \href
  {https://link.springer.com/content/pdf/10.1007/BFb0090557.pdf} {\emph
  {\bibinfo {booktitle} {Algebra Carbondale 1980: Lie Algebras, Group Theory,
  and Partially Ordered Algebraic Structures Proceedings of the Southern
  Illinois Algebra Conference, Carbondale, April 18 and 19, 1980}}}\ (\bibinfo
  {organization} {Springer},\ \bibinfo {year} {2006})\ pp.\ \bibinfo {pages}
  {63--83}\BibitemShut {NoStop}%
\bibitem [{\citenamefont {Djokovi{\'c}}(1982)}]{Djokovic1982}%
  \BibitemOpen
  \bibfield  {author} {\bibinfo {author} {\bibfnamefont {Dragomir~{\v{Z}}}\
  \bibnamefont {Djokovi{\'c}}},\ }\bibfield  {title} {\enquote {\bibinfo
  {title} {Closures of conjugacy classes in classical real linear lie groups.
  ii},}\ }\href {\doibase 10.1090/S0002-9947-1982-0642339-4} {\bibfield
  {journal} {\bibinfo  {journal} {Transactions of the American Mathematical
  Society}\ }\textbf {\bibinfo {volume} {270}},\ \bibinfo {pages} {217--252}
  (\bibinfo {year} {1982})}\BibitemShut {NoStop}%
\bibitem [{\citenamefont {Collingwood}\ and\ \citenamefont
  {McGovern}(2017)}]{Collingwood2017}%
  \BibitemOpen
  \bibfield  {author} {\bibinfo {author} {\bibfnamefont {David~H}\ \bibnamefont
  {Collingwood}}\ and\ \bibinfo {author} {\bibfnamefont {William~M}\
  \bibnamefont {McGovern}},\ }\href@noop {} {\emph {\bibinfo {title} {Nilpotent
  Orbits in Semisimple Lie Algebras}}}\ (\bibinfo  {publisher} {Routledge},\
  \bibinfo {year} {2017})\BibitemShut {NoStop}%
\bibitem [{\citenamefont {Trampus}(1966)}]{trampus1966}%
  \BibitemOpen
  \bibfield  {author} {\bibinfo {author} {\bibfnamefont {Anthony}\ \bibnamefont
  {Trampus}},\ }\bibfield  {title} {\enquote {\bibinfo {title} {A canonical
  basis for the matrix transformation x → ax + xb},}\ }\href {\doibase
  https://doi.org/10.1016/0022-247X(66)90024-2} {\bibfield  {journal} {\bibinfo
   {journal} {Journal of Mathematical Analysis and Applications}\ }\textbf
  {\bibinfo {volume} {14}},\ \bibinfo {pages} {242--252} (\bibinfo {year}
  {1966})}\BibitemShut {NoStop}%
\bibitem [{\citenamefont {Naghiloo}\ \emph {et~al.}(2019)\citenamefont
  {Naghiloo}, \citenamefont {Abbasi}, \citenamefont {Joglekar},\ and\
  \citenamefont {Murch}}]{naghiloo2019}%
  \BibitemOpen
  \bibfield  {author} {\bibinfo {author} {\bibfnamefont {M.}~\bibnamefont
  {Naghiloo}}, \bibinfo {author} {\bibfnamefont {M.}~\bibnamefont {Abbasi}},
  \bibinfo {author} {\bibfnamefont {Yogesh~N.}\ \bibnamefont {Joglekar}}, \
  and\ \bibinfo {author} {\bibfnamefont {K.~W.}\ \bibnamefont {Murch}},\
  }\bibfield  {title} {\enquote {\bibinfo {title} {Quantum state tomography
  across the exceptional point in a single dissipative qubit},}\ }\href
  {\doibase 10.1038/s41567-019-0652-z} {\bibfield  {journal} {\bibinfo
  {journal} {Nature Physics}\ }\textbf {\bibinfo {volume} {15}},\ \bibinfo
  {pages} {1232--1236} (\bibinfo {year} {2019})}\BibitemShut {NoStop}%
\bibitem [{\citenamefont {Chen}\ \emph {et~al.}(2021)\citenamefont {Chen},
  \citenamefont {Abbasi}, \citenamefont {Joglekar},\ and\ \citenamefont
  {Murch}}]{chen_quantum_2021}%
  \BibitemOpen
  \bibfield  {author} {\bibinfo {author} {\bibfnamefont {Weijian}\ \bibnamefont
  {Chen}}, \bibinfo {author} {\bibfnamefont {Maryam}\ \bibnamefont {Abbasi}},
  \bibinfo {author} {\bibfnamefont {Yogesh~N.}\ \bibnamefont {Joglekar}}, \
  and\ \bibinfo {author} {\bibfnamefont {Kater~W.}\ \bibnamefont {Murch}},\
  }\bibfield  {title} {\enquote {\bibinfo {title} {Quantum jumps in the
  non-{Hermitian} dynamics of a superconducting qubit},}\ }\href {\doibase
  10.1103/PhysRevLett.127.140504} {\bibfield  {journal} {\bibinfo  {journal}
  {Phys. Rev. Lett.}\ }\textbf {\bibinfo {volume} {127}},\ \bibinfo {pages}
  {140504} (\bibinfo {year} {2021})}\BibitemShut {NoStop}%
\bibitem [{\citenamefont {Bian}\ \emph {et~al.}(2020)\citenamefont {Bian},
  \citenamefont {Xiao}, \citenamefont {Wang}, \citenamefont {Zhan},
  \citenamefont {Onanga}, \citenamefont {Ruzicka}, \citenamefont {Yi},
  \citenamefont {Joglekar},\ and\ \citenamefont {Xue}}]{bian2020conserved}%
  \BibitemOpen
  \bibfield  {author} {\bibinfo {author} {\bibfnamefont {Zhihao}\ \bibnamefont
  {Bian}}, \bibinfo {author} {\bibfnamefont {Lei}\ \bibnamefont {Xiao}},
  \bibinfo {author} {\bibfnamefont {Kunkun}\ \bibnamefont {Wang}}, \bibinfo
  {author} {\bibfnamefont {Xiang}\ \bibnamefont {Zhan}}, \bibinfo {author}
  {\bibfnamefont {Franck~Assogba}\ \bibnamefont {Onanga}}, \bibinfo {author}
  {\bibfnamefont {Frantisek}\ \bibnamefont {Ruzicka}}, \bibinfo {author}
  {\bibfnamefont {Wei}\ \bibnamefont {Yi}}, \bibinfo {author} {\bibfnamefont
  {Yogesh~N.}\ \bibnamefont {Joglekar}}, \ and\ \bibinfo {author}
  {\bibfnamefont {Peng}\ \bibnamefont {Xue}},\ }\bibfield  {title} {\enquote
  {\bibinfo {title} {Conserved quantities in parity-time symmetric systems},}\
  }\href {\doibase 10.1103/PhysRevResearch.2.022039} {\bibfield  {journal}
  {\bibinfo  {journal} {Phys. Rev. Res.}\ }\textbf {\bibinfo {volume} {2}},\
  \bibinfo {pages} {022039} (\bibinfo {year} {2020})}\BibitemShut {NoStop}%
\bibitem [{\citenamefont {Agarwal}\ \emph {et~al.}(2022)\citenamefont
  {Agarwal}, \citenamefont {Muldoon},\ and\ \citenamefont
  {Joglekar}}]{agarwal2022conserved}%
  \BibitemOpen
  \bibfield  {author} {\bibinfo {author} {\bibfnamefont {Kaustubh~S.}\
  \bibnamefont {Agarwal}}, \bibinfo {author} {\bibfnamefont {Jacob}\
  \bibnamefont {Muldoon}}, \ and\ \bibinfo {author} {\bibfnamefont {Yogesh~N.}\
  \bibnamefont {Joglekar}},\ }\bibfield  {title} {\enquote {\bibinfo {title}
  {Conserved quantities in non-hermitian systems via vectorization method},}\
  }\href {\doibase 10.14311/AP.2022.62.0001} {\bibfield  {journal} {\bibinfo
  {journal} {Acta Polytechnica}\ }\textbf {\bibinfo {volume} {62}},\ \bibinfo
  {pages} {1–7} (\bibinfo {year} {2022})}\BibitemShut {NoStop}%
\bibitem [{\citenamefont {Horn}\ and\ \citenamefont
  {Johnson}(2012)}]{horn2012matrix}%
  \BibitemOpen
  \bibfield  {author} {\bibinfo {author} {\bibfnamefont {Roger~A}\ \bibnamefont
  {Horn}}\ and\ \bibinfo {author} {\bibfnamefont {Charles~R}\ \bibnamefont
  {Johnson}},\ }\href@noop {} {\emph {\bibinfo {title} {Matrix analysis}}}\
  (\bibinfo  {publisher} {Cambridge university press},\ \bibinfo {year}
  {2012})\BibitemShut {NoStop}%
\end{thebibliography}%
\end{document}